\documentclass[12pt,aip,pof,graphicx,amsmath,amssymb]{revtex4-1}
\usepackage{wasysym}
\usepackage{ifsym}
\usepackage{bm}
\usepackage{graphicx}
\usepackage{float}
\usepackage{latexsym}
\usepackage{color}
\usepackage{soul}
\usepackage[usenames,dvipsnames]{xcolor}
\usepackage{hyperref}
\hypersetup{
    colorlinks,%
    citecolor=blue,%
    filecolor=blue,%
    linkcolor=blue,%
    urlcolor=blue
} 

\begin{document}

\preprint{}

\title{Strong anisotropy in quasi-static MHD turbulence for high interaction parameters} 

\author{K. Sandeep Reddy}
\email{ksreddy@iitk.ac.in}
\affiliation{Department of Mechanical Engineering, Indian Institute of Technology Kanpur, India 208016}

\author{Mahendra K. Verma}
\affiliation{Department of Physics, Indian Institute of Technology Kanpur, India 208016}


\date{\today}

\begin{abstract}
We simulate forced quasi-static magnetohydrodynamic turbulence and investigate the anisotropy, energy spectrum, and energy flux of the flow, specially for large interaction parameters $(N)$.  We show that the angular dependence of the energy spectrum is well quantified using Legendre polynomials.  For large $N$, the energy spectrum is exponential.  Our direct computation of energy flux reveals an inverse cascade of energy at low wavenumbers, similar to that in two-dimensional turbulence.  We observe the flow be two-dimensional (2D) for moderate $N$ ($N \sim 20$), and two-dimensional three-component (2D-3C) type for $N \ge 27$.  In our forced simulation, the transition from 2D to 2D-3C occurs at higher value of $N$ than Favier \textit{et al.,} [B. Favier, F. S. Godeferd, C. Cambon, A. Delache, ``On the two-dimensionalization of quasistatic magnetohydrodynamic turbulence," Phys. Fluids {\bf 22}, 075104 (2010)] who employ decaying simulations.
\end{abstract}


\maketitle 

\section{Introduction} \label{sec:Introduction}


Magnetohydrodynamics (MHD) is used for studying flows of conducting fluids and plasmas.  MHD flows involving plasma are observed in the sun, stars, solar flares, Tokamac, etc., while those involving conducting fluids are found in the core of the Earth, liquid-metal flows in industries, and laboratory experiments. \cite{Alemany:JMec1979,Kolesnikov:FD1974,Monchaux:PRL2007}   A major application of liquid metal MHD is in International Thermonuclear Experimental Reactor (ITER), where liquid metals are used as a heat exchanger.    In ITER, a strong external magnetic field affects the properties of the flow.  Thus, a study of the effects of the external magnetic field on the flow is critical for the design of heat exchangers.  An idealized version of the flow, called quasi-static approximation,\cite{Roberts:book,Moffatt:JFM1967,Knaepen:ARFM2008} has vanishing magnetic Reynolds number (${{Rm}} \rightarrow 0$) and magnetic Prandtl number (${{Pm}} \rightarrow 0$). ${{Rm}}$ and ${{Pm}}$ for most of the industrial flows involving liquid metals fall in this regime.  In this paper, we investigate properties of MHD flows in the quasi-static limit. 

In liquid metal MHD, a non-dimensional number called  ``interaction parameter''
\begin{equation}
N = \dfrac{\sigma {B_0}^2 L}{\rho u^\prime},
\end{equation}
plays an important role in determining flow properties.   Here $B_0$ is the external magnetic field, $L$ is the integral length scale, $\rho$, $\sigma$ are the density and conductivity of the fluid respectively, and $u^\prime$ is the rms value of the velocity fluctuations.  In our paper we calculate $N$ using $u^\prime$ and $L$ of the steady-state flow after application of an external magnetic field.  This is in contrast to earlier work where $N$ is measured using $u^\prime$ and $L$ at the instant when the magnetic field is applied (to be described in Sec.~\ref{sec:simulation procedure}).
Moffatt\cite{Moffatt:JFM1967} studied quasi-static MHD in the asymptotic limit of $N\gg1$, where the flow becomes two-dimensional. Sommeria and Moreau\cite{Sommeria:JFM1982} proposed that the diffusion of momentum in the direction of magnetic field elongates the vortical structures along the magnetic field.
 Alemany \textit{et al.}\cite{Alemany:JMec1979} and Kolesnikov and Tsinober\cite{Kolesnikov:FD1974}  studied quasi-static MHD by experimenting with mercury under a strong external field. They observed  that the kinetic energy spectrum scales as $k^{-3}$ for $N$ around unity. Their results showed experimental evidence of two-dimensional flow. 
 
 Branover \textit{et al.}\cite{Branover:PTR1994} performed experiments on mercury under a strong transverse magnetic field. In their experiments they observed different  energy spectra ($k^{-5/3}$, $k^{-7/3}$, $k^{-3}$, and $k^{-11/3}$) as a function of $N$. Branover \textit{et al.}\cite{Branover:book} explained this behavior based on helical nature of the flow. Eckert \textit{et al.}\cite{Eckert:HFF2001} performed experiments  in a channel under a strong external magnetic field with liquid sodium as a fluid. They showed that the exponent $\alpha$ of the energy spectrum $k^{\alpha}$ decreases with increasing $N$. Klein and Poth{\'e}rat\cite{Klein:PRL2010} performed experiments  on a wall bounded geometry and studied the transition from two-dimensional flow to three-dimensional flow. They observed that the eddy currents in the boundary layer and in the core were responsible for three-dimensionalization of the flow.  Poth{\'e}rat\cite{Potherat:EPL2012} proposed that ``barrel effect" is  responsible for transforming a quasi 2D flow to 3D flow in wall bounded geometries.  Note that experiments involving liquid metals (primarily mercury and liquid sodium) have major practical difficulties in their implementation as well as in visualization. Numerical simulations  play an important complementary role in this field, and enable us to probe the flow profiles inside the box, specially for idealised geometries.

For studying the properties of bulk flow, it is customary to employ direct numerical simulation (DNS), mostly using pseudospectral method on a box geometry.  Hossain\cite{Hossain:POP1991} performed forced DNS and reported that for low interaction parameter ($N=0.1$), the flow is three-dimensional and it exhibits a forward cascade of energy to higher wavenumbers. However at $N = 10$, the flow is quasi two-dimensional with an inverse cascade of energy to lower wavenumbers. Zikanov and Thess\cite{Zikanov:JFM1998} performed forced  DNS and studied anisotropy in the velocity field. They observed that the flow remains three-dimensional and turbulent for low interaction parameters ($N = 0.1$), quasi-two-dimensional with sporadic three-dimensional bursts for moderate interaction parameters ($N=0.4$), and fully two-dimensional for high interaction parameters ($N=10$). Schumann\cite{Schumann:JFM1976} simulated decaying quasi-static MHD and observed that for $N\geq50$, the flow is quasi two-dimensional, with a reduced energy transfer for the velocity components perpendicular to the external magnetic field, and a higher energy transfer for the parallel velocity component.  Knaepen \textit{et al.}\cite{Knaepen:JFM2004} compared numerical results of quasi-static MHD with those with moderate magnetic Reynolds number and found significant similarities. Boeck \textit{et al.}\cite{Boeck:PRL2008} performed numerical simulations in a wall bounded flow with transverse magnetic field, and observed  large-scale Intermittency, where a 2D flow suddenly transforms to a 3D flow.

Burattini \textit{et al.}\cite{Burattini:PD2008} studied nonlinear energy transfers and showed that the energy flux is both radial and angular.  They also studied the  anisotropic distribution of energy as a function of the interaction parameter.  Burattini \textit{et al.}\cite{Burattini:PF2008} computed 1D and 3D spectra from DNS.   In the simulations presented in this paper we also observe that the exponent of energy spectrum decreases with $N$.  However the spectrum is exponential for very large $N$.   Using analytical arguments, Verma\cite{Verma:ARXIV2013a} showed that the increase in the spectral exponent with the interaction parameter is related to the variable energy flux, which occurs due to the Joule dissipation.  

Vorobev \textit{et al.}\cite{Vorobev:POF2005} quantified the flow anisotropy using $k$-dependent energy spectrum. For $N=5$, they observed that $E_{\perp}(k)/E_{\parallel}(k) > 1$ at low wavenumbers ($k$), and $E_{\perp}(k)/E_{\parallel}(k) < 1$ at higher wavenumbers.    In a recent work, Favier \textit{et al.}\cite{Favier:PF2010b}  performed decaying simulation for $N=1-5$ and showed that the quasi-static MHD flow is more complex than two-dimensional flow.  They showed that the flow is better described by two-dimensional-three-component (2D-3C); the horizontal flow (perpendicular to the mean field) resembles two-dimensional turbulence, whereas the parallel component has similarities with a passive scalar advected by the 2D turbulence.  They argue in favor of an inverse cascade for the horizontal velocity, but for a forward cascade for the parallel component.    Favier \textit{et al.}\cite{Favier:JFM2011} also applied  eddy-damped quasi-normal Markovian (EDQNM) approximation to the quasi-static MHD, and observed that the model predictions are in good agreement with their numerical results.   

\begin{figure}[htbp]
 \begin{center}
 \includegraphics[scale=0.35]{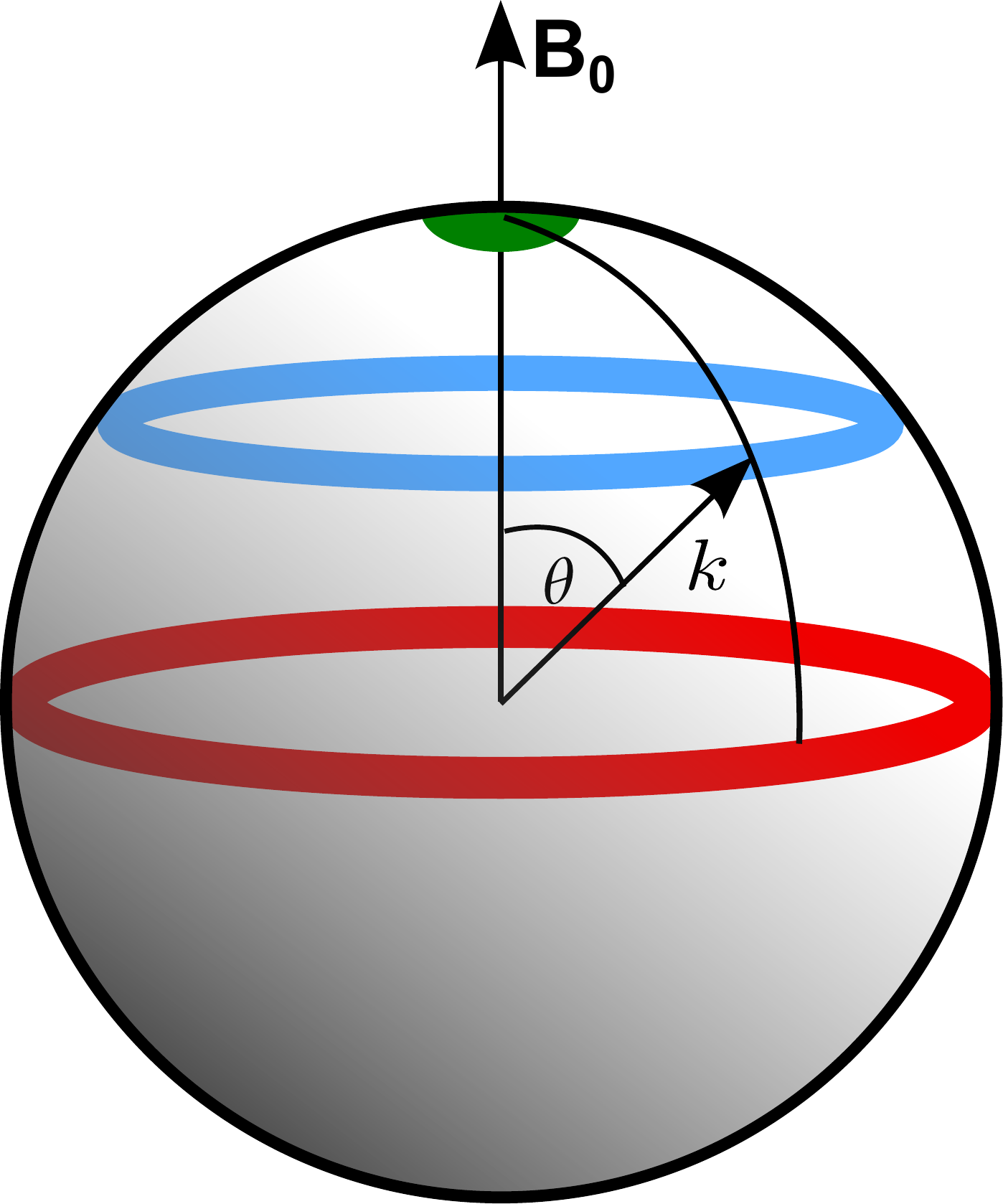}
 \end{center}
\caption{Figure illustrating ring decomposition in spectral space.}  
 \label{fig:ring_decomposition}
 \end{figure}

As described above, most of  the earlier numerical studies on quasi-static MHD have $N\le 10$.  However, some of the critical applications have much larger interaction parameters.  For example, the interaction parameter in ITER can be estimated to be ${Ha}^2/{Re} \approx 10^5$ using Hartmann number ${Ha} = 10^4$ and Reynolds number ${Re}=10^3 $.\cite{Buhler:FED2013,Smolentsev:FED2010}  Hartmann number is the ratio of Lorentz force and viscous force,  defined as $Ha = BL \sqrt{\sigma/\rho \nu}$.  We study forced quasi-static MHD for large $N$ ($\sim 200$).  We will show later in our discussion that the forced and decaying simulations exhibit some similarities and some dissimilarities.

In this paper, we perform numerical simulations of forced quasi-static MHD for $N = 0 - 220$.  Using the numerical data we study the flow anisotropy, energy spectrum, and energy flux.  The energy spectrum for very large $N$ is shown to be exponential, which differs from the power-law spectrum reported in earlier works.   We quantify the flow anisotropy by computing ring spectrum\cite{Teaca:PRE2009,Burattini:PD2008} for various $N$'s (see Fig.~\ref{fig:ring_decomposition}).  Further, we use group-theoretic basis functions like Legendre polynomials to describe the angular dependence of the energy.  Our work has certain similarities with those of Favier \textit{et al.,}\cite{Favier:PF2010b} but there are distinct differences.  We show that the flow is two-dimensional three-component (2D-3C) only for very large $N$ (e.g., for $N\ge 27$), but it is two-dimensional for moderate  $N$ (e.g., for $N=18$).  This result differs from that of  Favier~\textit{et al.}\cite{Favier:PF2010b} who report 2D-3C flow for $N=5$.  The difference arises due to forcing applied in our flow (contrary to the decaying simulations of Favier \textit{et al.}\cite{Favier:PF2010b}).  We will contrast our results with the aforementioned earlier work in later part of the paper.

  For low $Rm$ flows the non-local character of the Lorentz force makes the flow properties in periodic box simulations somewhat different from the wall bounded flows.  Yet, simulations with periodic box provide useful insights into the physics of the bulk flow. Kolmogorov\cite{Kolmogorov:DANS1941a} provided a theory of homogeneous and isotropic turbulence that quantifies the properties of the small-scale turbulence reasonably well.  Many researchers have  undertaken similar studies in other fields of turbulence, e.g., shear, MHD, scalar, quasi-static MHD, convective, rotating, stratified, etc.  Undoubtedly, anisotropy and walls play major part in the flow dynamics.\cite{Hunt:JFM1965,Grossmann:JFM2000} For example, in convective turbulence, walls induce a completely new branch in the entropy spectrum.\cite{Mishra:PRE2010} Hartmann profile provides an exact solution to the laminar solution of quasi-static MHD, while studies with periodic boundary conditions attempt to study the nonlinear effects in the bulk.  Future experimental and realistic numerical simulations would attempt to combine the effects of the bulk and boundary layer in the spirit of Grossmann and Lohse.\citep{Grossmann:JFM2000} Our study is motivated towards that attempt.

The paper is structured as follows. We introduce the governing equations in Sec.~\ref{sec:equations}. Simulation procedure is described in Sec.~\ref{sec:simulation procedure}.  Flow anisotropy and visualization are described in Sec.~\ref{sec:anisotropy}. Angular distribution of the kinetic energy and its representation using Legendre polynomials are described in Sec.~\ref{sec:Legendre}. In Sec.~\ref{sec:spectrum of KE}, we describe the spectrum and flux of the kinetic energy. Finally, we summarize the results in Sec.~\ref{sec:conclusion}.

\section{Governing Equations} \label{sec:equations}

Governing equations of liquid metal MHD under quasi-static approximation\cite{Roberts:book,Knaepen:ARFM2008} are 
\begin{eqnarray}
\dfrac{\partial{\bf u}}{\partial t} + ({\bf u}\cdot\nabla){\bf u} &=& -\nabla{(p/\rho)} - \dfrac{\sigma {B_0}^2}{\rho} \dfrac{1}{\nabla^2} \dfrac{\partial^2{\bf u}}{\partial z^2} + \nu\nabla^2 {\bf u} + {\bf f}, \label{eq:NS}\\
\nabla \cdot {\bf u} & = & 0, \label{eq:continuity} 
\end{eqnarray}
where {\bf u} is the velocity field, $\mathbf B_0 = B_0\hat{z}$ is the constant external magnetic field, $p$ is the thermal pressure, $\rho$ is the density, $\nu$ is the kinematic viscosity, $\sigma$ is the conductivity, and ${\bf f}$ is the forcing field.

 We nondimensionalize Eqs.~(\ref{eq:NS},\ref{eq:continuity})  using the characteristic velocity $U_0$ as the velocity scale, the length of the box $L_0$ as the length scale, and $L_0/U_0$ as the time scale, that yields
\begin{eqnarray}
\dfrac{\partial{\bf U}}{\partial T} + ({\bf U}\cdot\nabla'){\bf U} &=& -\nabla'{P} - B^{\prime 2}_0 \dfrac{1}{\nabla^{\prime2}} \dfrac{\partial^2{\bf U}}{\partial Z^2} + \nu^\prime \nabla^{\prime 2} {\bf U} + {\bf f^\prime}, \label{eq:NS2}\\
\nabla^\prime \cdot {\bf U} & = & 0, \label{eq:continuity2} 
\end{eqnarray}
\noindent
where non-dimensional variables $\mathbf U = \mathbf u/U_0$, $ \nabla' = L_0 \nabla $, $T = t(U_0/L_0)$, 
$B_0^{\prime 2}  = \sigma B_0^2 L_0 /(\rho U_0)$ and $\nu^\prime=\nu/(U_0 L_0)$.  The above equations when transformed in the Fourier space\cite{Schumann:JFM1976,Zikanov:JFM1998,Knaepen:JFM2004} yields
\begin{eqnarray}
\dfrac{\partial{\hat{U}_i(\bf{k})}}{\partial T} + ik_j \sum \hat{U}_j({\bf q}) \hat{U}_i({\bf{k}-\bf{q}})  & = & - ik_i \hat{P}({\bf k}) - {B'_0}^2{\mathrm{cos^2}}(\theta)\hat{U}_i({\bf k}) - \nu' k^2 \hat{U}_i({\bf k})+\hat{f}_i({\bf k}), \\
k_i  \hat{U}_i(\mathbf{k}) & = & 0,
\end{eqnarray}
where $\hat{U}_i(\mathbf{k})$ is the Fourier transform of the velocity field, and $\theta$ is the angle between wavenumber vector ${\bf k}$ and the external magnetic field $\mathbf B_0$.

An important non-dimensional number in quasi-static MHD is the interaction parameter ($N$), which is defined as the ratio of Lorentz force term and the nonlinear term calculated as 
\begin{equation}
N = \dfrac{B_0^{\prime^2} L}{U^\prime},
\end{equation}
\noindent
where $L$ is the non-dimensional integral length scale, and $U'$ is rms of the fluctuating velocity. The total energy of the system and the integral length scale are defined as\cite{Vorobev:POF2005,Burattini:PD2008}

\begin{equation}
E =  \int_0^{\infty}E(k)dk = \frac{3}{2}U'^2,
\end{equation}
\begin{equation}
L = \dfrac{\pi}{(2{U'}^2)} \int_0^{\infty}(E(k)/k)dk,
\end{equation}
respectively. The eddy turnover time is defined as $\tau=L/U'$.
The equation for the evolution of kinetic energy in the Fourier space is\cite{Burattini:PF2008}
\begin{equation}
\dfrac{\partial{E({\bf k})}}{\partial t} = T({\bf k}) - 2 {{B'_0}^2} E({\bf k}){\mathrm{cos^2}}(\theta) - 2\nu' k^2 E({\bf k}) + F({\bf k}) ,
\label{eq:Ek}
\end{equation}
where  $E({\bf k}) = |\hat{U}({\bf k})|^2/2$ is the energy of the Fourier mode ${\bf k}$, $T({\bf k})$ is the rate of nonlinear energy transfer to the mode, and $F({\bf k})$ is the contribution of forcing to the energy equation.  The other two terms are the Joule dissipation rate $\epsilon_J$, and the viscous dissipation rate $\epsilon_\nu$ respectively, i.e., 
\begin{equation}
\epsilon_J = 2{B'_0}^2 \sum_{\bf{k}} E(\bf{k})\mathrm{cos^2}(\theta),
\end{equation}
\begin{equation}
\epsilon_\nu = 2 \nu' \sum_{\bf{k}} k^2E(\bf{k}). 
\end{equation}
We can interpret $\epsilon_J$ as the energy transfer from the velocity field to the magnetic field, which is instantaneously dissipated due to  infinite resistivity.  Also note that  the Joule dissipation is active at all scales unlike the viscous dissipation rate that dominates at small scales.   

The Reynolds number, which is the ratio of the nonlinear term and the viscous term, is defined as 
\begin{equation}
{Re} = \frac{U'L}{\nu^\prime}.
\end{equation}

\section{Simulation Method} \label{sec:simulation procedure}
We numerically solve Eqs.~(\ref{eq:NS2},\ref{eq:continuity2}) using pseudo-spectral method\cite{Canuto:book, Boyd:book} in a cubical box with periodic boundary condition on all sides.  We use the fourth-order Runge-Kutta method for time stepping, and the Courant-Friedrichs-Lewy (CFL) condition for calculating time step $\Delta t$. We also apply $3/2$ rule  for dealiasing. The grid resolution of our simulations is $256^3$, which is sufficient for the parameters explored in our simulations. All the simulations have been performed using a pseudo-spectral code \emph{Tarang}.\cite{Verma:Pramana2013} The value of non-dimensional $\nu^\prime = 0.00036$ is fixed, and non-dimensional $B^\prime_0$ is varied (see Table~\ref{tab:parameters}).

Our simulations reach a statistically steady state (approximate constant energy) after several eddy turnovers. We compute energy spectra and related quantities for the steady states.   First we perform a simulation for $N=0$ with the following initial energy spectrum \cite{Pope:book}:
\begin{equation}
E(k) = C\epsilon^{2/3}k^{-5/3}f_L(kL)f_{\eta}(k\eta),
\end{equation} 
The Fourier modes are assigned random phases.  We choose  $C = 1.0$,  $\epsilon = 1.0$, and
\begin{eqnarray}
f_L(kL) &=& \left(\frac{kL}{[(kL)^2+c_L]^{1/2}}\right)^{5/3+p_0},\\
f_{\eta}(k\eta) &=& exp(-\beta k \eta),
\end{eqnarray}
where $c_L=1.5$, $p_0=2$,  $\beta = 5.2$, and $\eta$ is the Kolmogorov length scale.   For forcing, we use a scheme similar to that proposed by Vorobev {\textit {et al.}}\cite{Vorobev:POF2005} and Burattini {\textit {et al.,}}\cite{Burattini:PF2008} and  apply the following forcing function within a wavenumber shell $1.0 \le k \le 3.0$:
\begin{eqnarray}
{\bf {\hat f}(k)} &=& \gamma({\bf k}){\bf {\hat U}(k)},\\
\gamma({\bf k}) &=& \frac{\epsilon_{in}}{n_{f}(\hat{\bf U}({\bf k})\hat{\bf U}^*({\bf k}))},
\end{eqnarray}
where $n_f$ is number of modes in the shell $1.0 \le k \le 3.0$, and $\epsilon_{in} = 0.1$ is the input energy supply rate ($dE/dt$). The final state of the above (hydrodynamic) run is used as the initial condition for the simulations with non-zero $N$. We carry out our simulations (for non-zero $N$'s) till another statistically steady state is reached.  We compute the value of $N$ using ${U}'$ and $L$ of the steady state data.  This notation differs from the procedure adopted in earlier work where $N$  is calculated using $U'$ and $L$ computed at beginning of the simulation, i.e., at an instant just before applying external magnetic field; we denote this interaction parameter as $N_0$ in Table~\ref{tab:parameters} and in subsequent discussion.  The steady state of $ N =27$ is chosen as an initial condition for the simulations with $N = 130$ and $220$ in order to reach steady states quickly. The value of $k_{\mathrm{max}}\eta$ is greater than $1.4$ in all our simulations, where $\eta$ is the Kolmogorov length scale, and $k_{\mathrm{max}}$ is the maximum wavenumber attained in DNS for a particular grid size. By this criterion, the smallest grid size in our simulation is smaller than Kolmogorov length scale, and all the flow scales are fully resolved.\cite{Favier:PF2010b,Jimenez:JFM1993}

\begin{figure}[htbp]
 \begin{center}
 \includegraphics[scale=0.55]{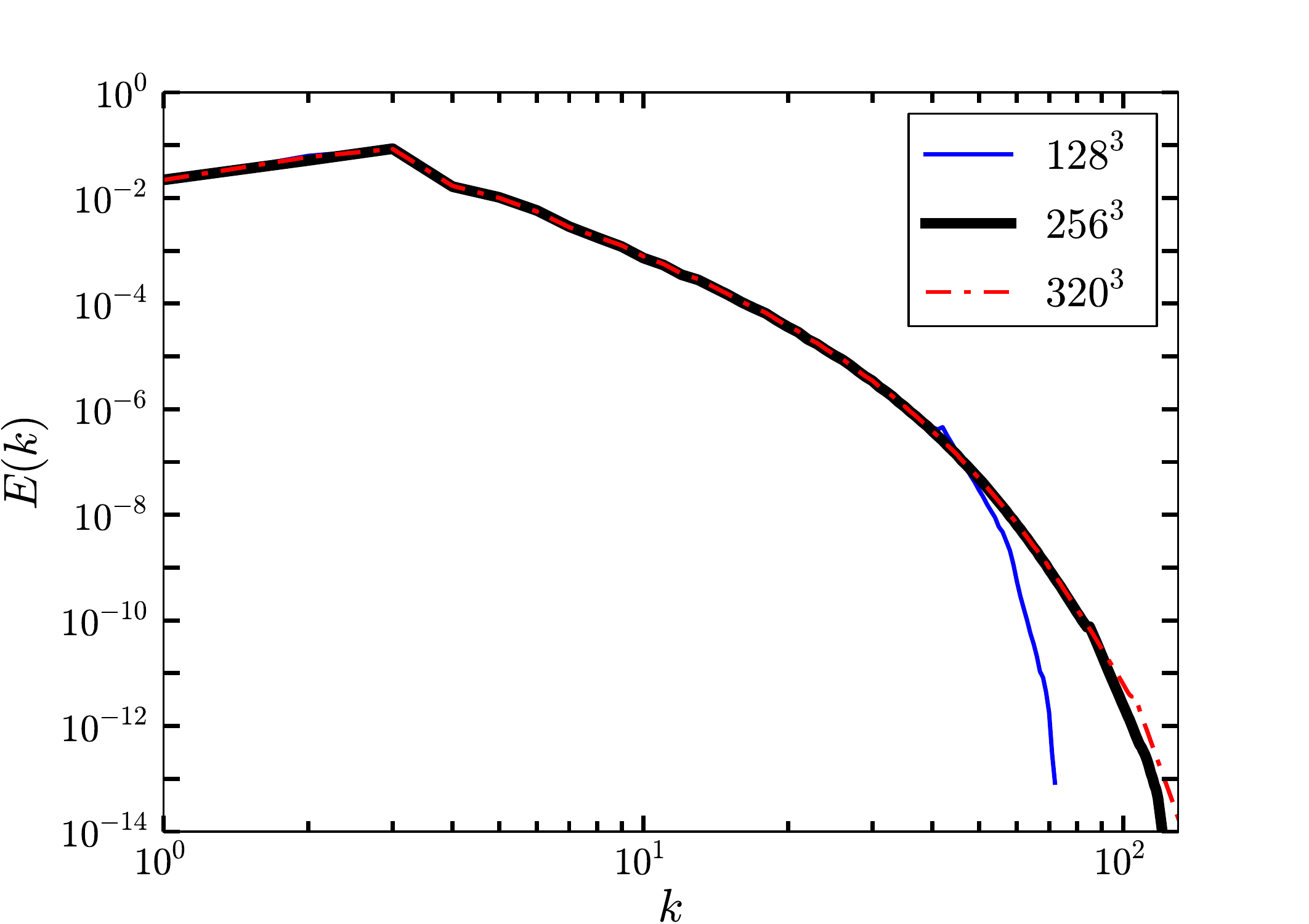}
 \end{center}
 \caption{Energy spectrum for grids $128^3$, $256^3$, and $320^3$ for $N = 5.5$. The small scales are well resolved for grids $256^3$ and $320^3$.}  
 \label{fig:grid}
 \end{figure}

We performed grid independence test for $N=5.5$ using $128^3$, $256^3$, and $320^3$ grids.  We observe that $256^3$ and $320^3$ grids have similar energy spectra,  and they resolve the small scales better than $128^3$ grid (see Fig.~\ref{fig:grid}). The integral length scale $L$ obtained for $128^3$, $256^3$, and $320^3$ grids are 0.17, 0.15, and 0.15 respectively. We find $k_\mathrm{max}\eta=1.2$ for $128^3$ grid, and  $k_\mathrm{max}\eta \approx 2.1$ for the larger grids. We observe that the energy spectrum, integral length scale, total energy, and $k_{\mathrm{max}}\eta$ are the same  for the grid sizes of  $256^3$ to $320^3$. Hence, the grid size $256^3$, chosen for all our simulations, is sufficient for our study.

\begin{table}
\caption{Parameters of the simulation: the constant external magnetic field $B^\prime_0$, rms velocity at steady state $U'$, the interaction parameter $N$ computed at steady state, the interaction parameter $N_0$ computed at the instant when external magnetic field is applied, the ratio of the Joule dissipation and viscous dissipation $\epsilon_J/\epsilon_{\nu}$, Reynolds number $Re$, the energy spectrum, $E_{\perp}/2E_{\parallel}$, and eddy turnover time $\tau$ based on the steady state, i.e., $\tau=L/U'$.} 

\begin{center}
\begin{tabular}{ p{1.0cm}  p{1.1cm}  p{1.0cm}  p{0.9cm} p{1.0cm}  p{1.2cm}  p{2.4cm}  p{1.9cm}  p{0.7cm}}
 \hline 
 $B_0^\prime$ 	 &~\ $U'$ 		&~\ $N$		&~\ $N_0$	& ~\ $\epsilon_J/\epsilon_{\nu}$	&~\ ${Re}$		&   spectrum	 & $E_{\perp}/2E_{\parallel}$  & $\tau$   \\
    
 \hline
  2.29		&~\ 0.39		&~\ 1.7		&~\ 1.0		&~\ 4.2			&~\ 130		& ~\ $k^{-3.2}$	& 1.1  & 0.32 \\
  3.60		&~\ 0.35		&~\ 5.5		&~\ 2.5		&~\ 9.7			&~\ 140		& ~\ $k^{-3.8}$	& 1.5	& 0.43 \\
  5.15		&~\ 0.39		&~\ 11	&~\ 5.0		&~\ 11			&~\ 170		& ~\ $k^{-4.0}$	& 4.5	& 0.39 \\
  6.26		&~\ 0.45		&~\ 14	&~\ 7.5		&~\ 11			&~\ 210		& ~\ $k^{-4.5}$	& 8.0	& 0.37 \\
  7.28		&~\ 0.51		&~\ 18	&~\ 10.0	&~\ 9.8			&~\ 240		& ~\ $k^{-4.7}$	& 16	& 0.33 \\
  10.23	   &~\ 0.65		&~\ 27	&~\ 20.0	&~\ 6.9			&~\ 300		& ~\ $k^{-4.7}$	& 1.6	& 0.26 \\
  25.1	   &~\ 0.86		&~\ 130	&~\ $-$		&~\ 4.1			&~\ 430		&~\ exp(-0.18$k$) & 3.0 & 0.21 \\
  32.6    &~\ 0.87		&~\ 220	&~\ $-$		&~\ 2.8			&~\ 440		&~\ exp(-0.18$k$)	& 1.7 & 0.21 \\ 			
 \hline
\end{tabular}
\end{center}
\label{tab:parameters}
\end{table}

\begin{figure}[htbp]
\begin{center}
\includegraphics[scale=0.64]{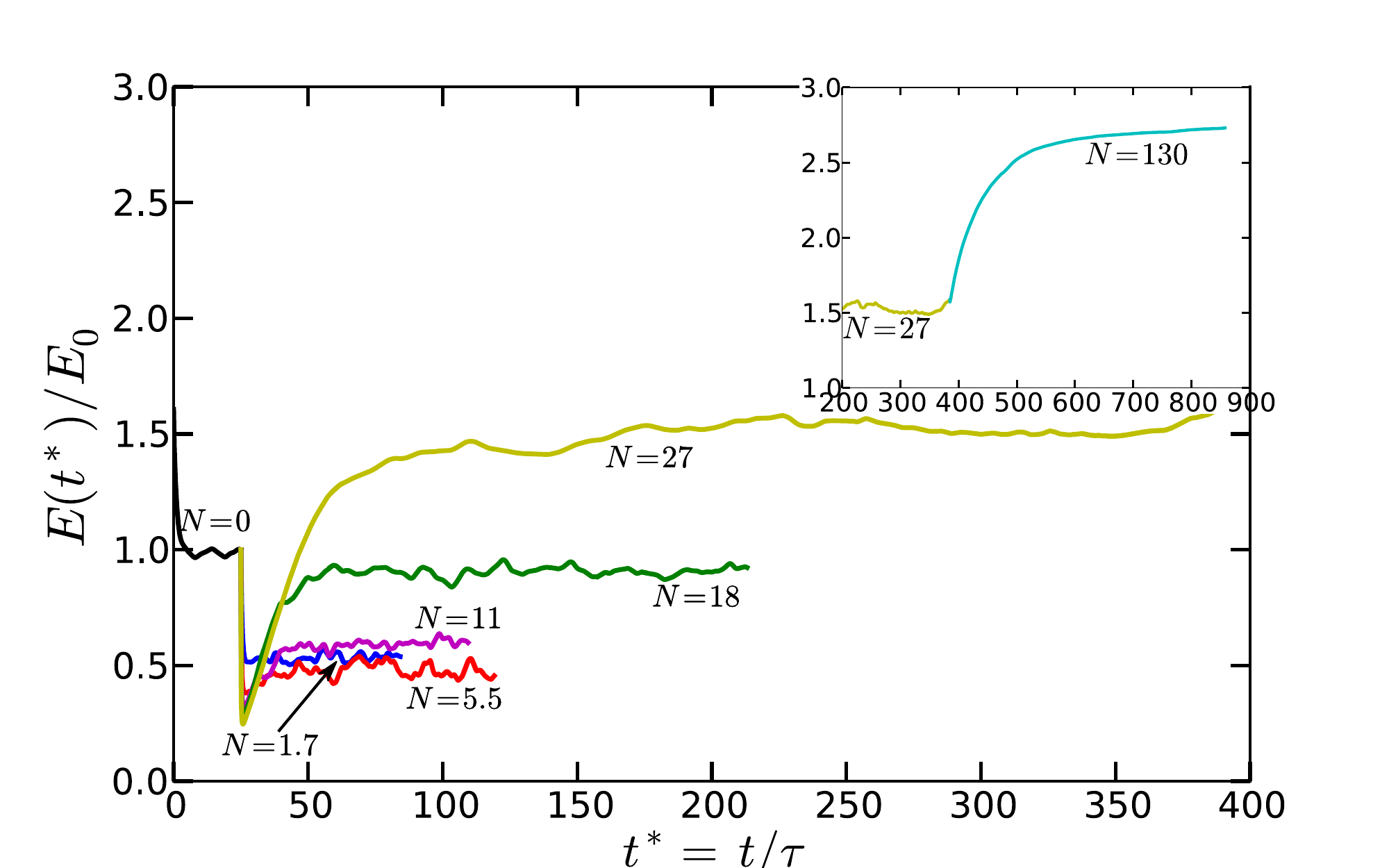}
\end{center}
\caption{Time evolution of normalised total energy $E(t)/E_0$ for different interaction parameters $N$; here $E_0$ is the energy at the final state of $N=0$ simulation. The energy drops immediately after the application of external magnetic field. After the dip, the energy increases and reaches a statistically steady state. For $N=27$, the energy at final steady state is greater than the $N =0$. The subfigure shows the time series for $N=130$. } 
\label{fig:global_energy}
\end{figure}

Figure~\ref{fig:global_energy} exhibits evolution of energy for different interaction parameters. The kinetic energy of the system decreases immediately after an external magnetic field is applied.  This is due to the well-known suppression of energy flux by the mean magnetic field.  After a dip, the total energy of the system reaches a new steady state. The asymptotic level of the total energy increases with $N$.  For $N=27$ and above, we observe that the the energy increases after a sharp decline, and then reach relatively higher energy levels. The two-dimensionalization of the flow suppresses the Joule dissipation due to $\cos^2\theta$ factor, and the level of energy  for a forced simulation increases with $N$ for a given energy supply rate.\cite{Kanaris:POF2013}
We point out that our simulations have been carried out up to 200 to 400 eddy turnover times, which is much larger than most of the earlier simulations.  

We performed our simulations for various sets of parameters ($B'_0$ or $N$).  The parameters of the simulations are shown in Table~\ref{tab:parameters}.   We will discuss the properties of the flow for these parameters in the subsequent sections.

\section{Anisotropy in Liquid Metal MHD} \label{sec:anisotropy}

\begin{figure}[htbp]
\begin{center}
\includegraphics[scale=0.64]{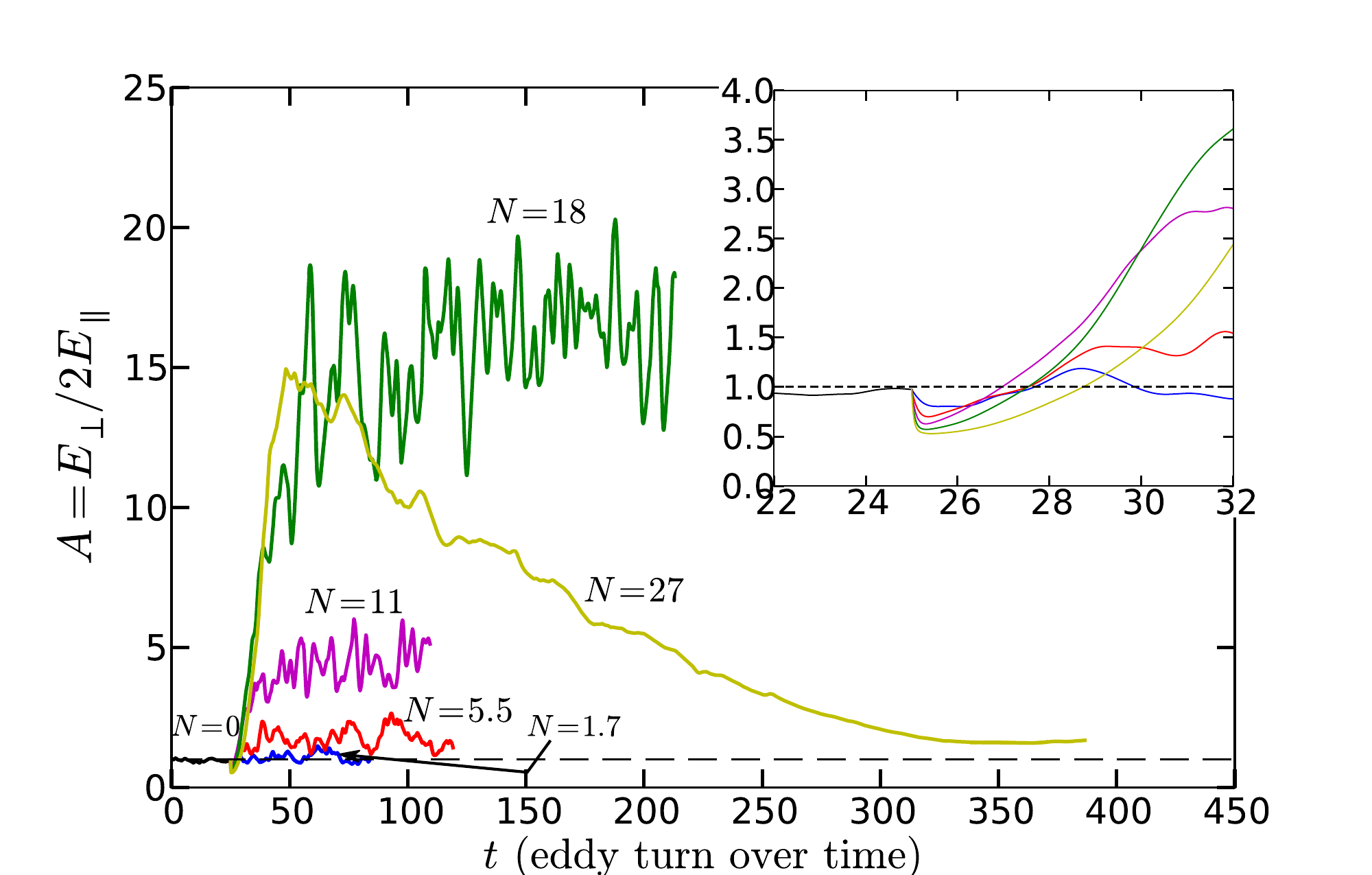}
\end{center}
\caption{Time evolution of $A=E_\perp/2 E_{\parallel}$ for different interaction parameters $N$. The subfigure shows the evolution of A at the early stages when the external magnetic field is applied.} 
\label{fig:time_eperp_epar}
\end{figure}

The flow is isotropic in the absence of external magnetic field.  But it becomes anisotropic when an external field is applied, with the degree of anisotropy increasing with strength of the external field or  $N$.  One of the quantitative measure of anisotropy is the ratio $A = E_{\perp}/2E_{\parallel}$, where $E_{\perp} = (u^2_x+u^2_y)/2$, and  $E_{\parallel} = u^2_z/2$.  Physically, $E_{\perp}$ and $E_{\parallel}$ denote the energy components perpendicular and parallel to the mean magnetic field respectively.
For isotropic flows, $A=1$ since all the components have approximately equal energy.  On the other hand, $A$ deviates from unity for anisotropic flows.  In Fig.~\ref{fig:time_eperp_epar}, we plot the evolution of $A$ as a function of time.  The ratio decreases in the beginning and then increases.  The asymptotic or steady-state values of $A$ for various $N$'s are listed in Table~\ref{tab:parameters}.  The trend clearly demonstrates an increase of anisotropy with the increase of $N$ till  $N = 18$, after which it drops suddenly.   It is interesting to contrast our results with those of Favier \textit{et al.}\cite{Favier:PF2010b,Favier:JFM2011}  for decaying simulations, according to which $A(t)$ is less than 1.5 for $N_0=5$.  Favier \textit{et al.}'s\cite{Favier:PF2010b,Favier:JFM2011} data shows an increasing trend for $A(t)$ at $t=t_\mathrm{max}=1.9$ of their simulation; it is possible that $A(t)$ may saturate at a higher value  at a later time even in the decaying simulation for $N_0=5$.

\begin{figure}[htbp]
\begin{center}
\includegraphics[scale=0.217]{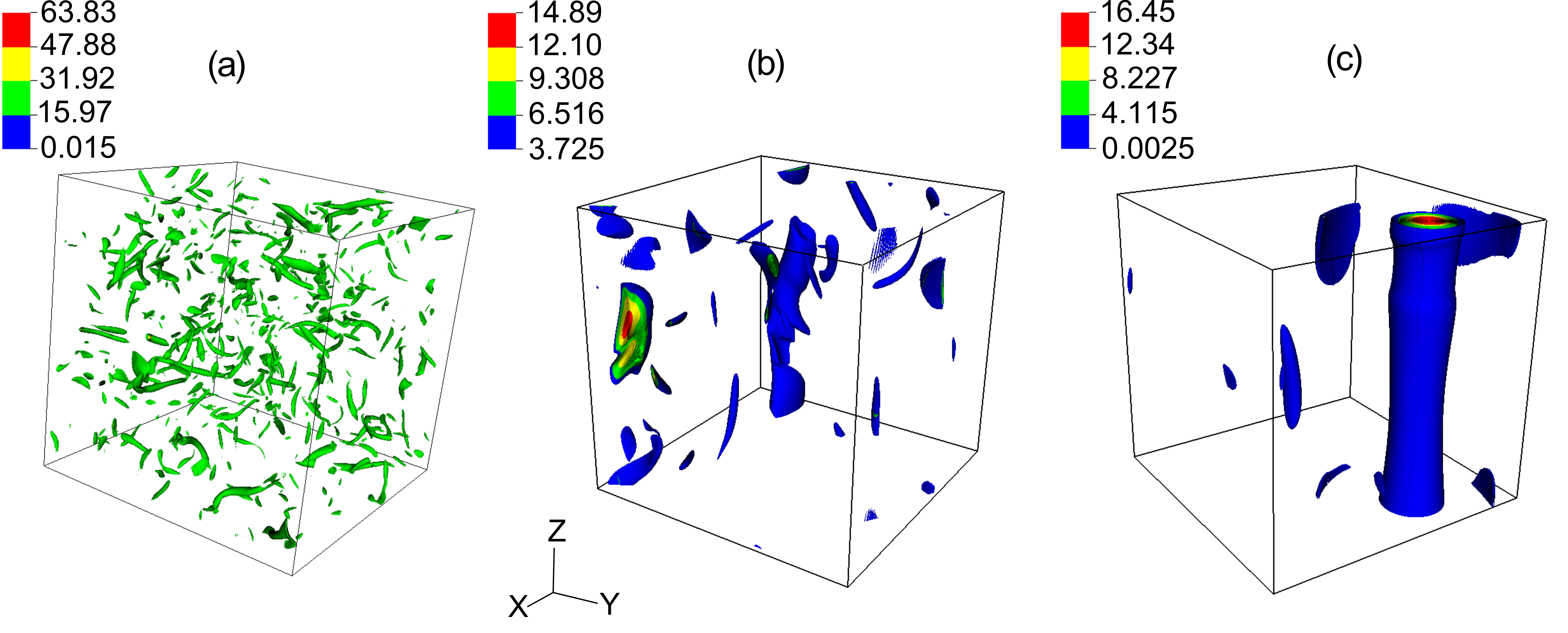}
\end{center}
\caption{Isosurfaces of vorticity for (a) $N = 0$, (b)  $N = 5.5$ and (c)  $N = 18$. The flow field becomes anisotropic for $N\ne 0$. For higher interaction parameters, a vortex tube is formed with its axis in the direction of the external magnetic field.}  
 \label{fig:isosurface_vorticity}
\end{figure}

\begin{figure}[htbp]
 \begin{center}
 \includegraphics[scale=0.33]{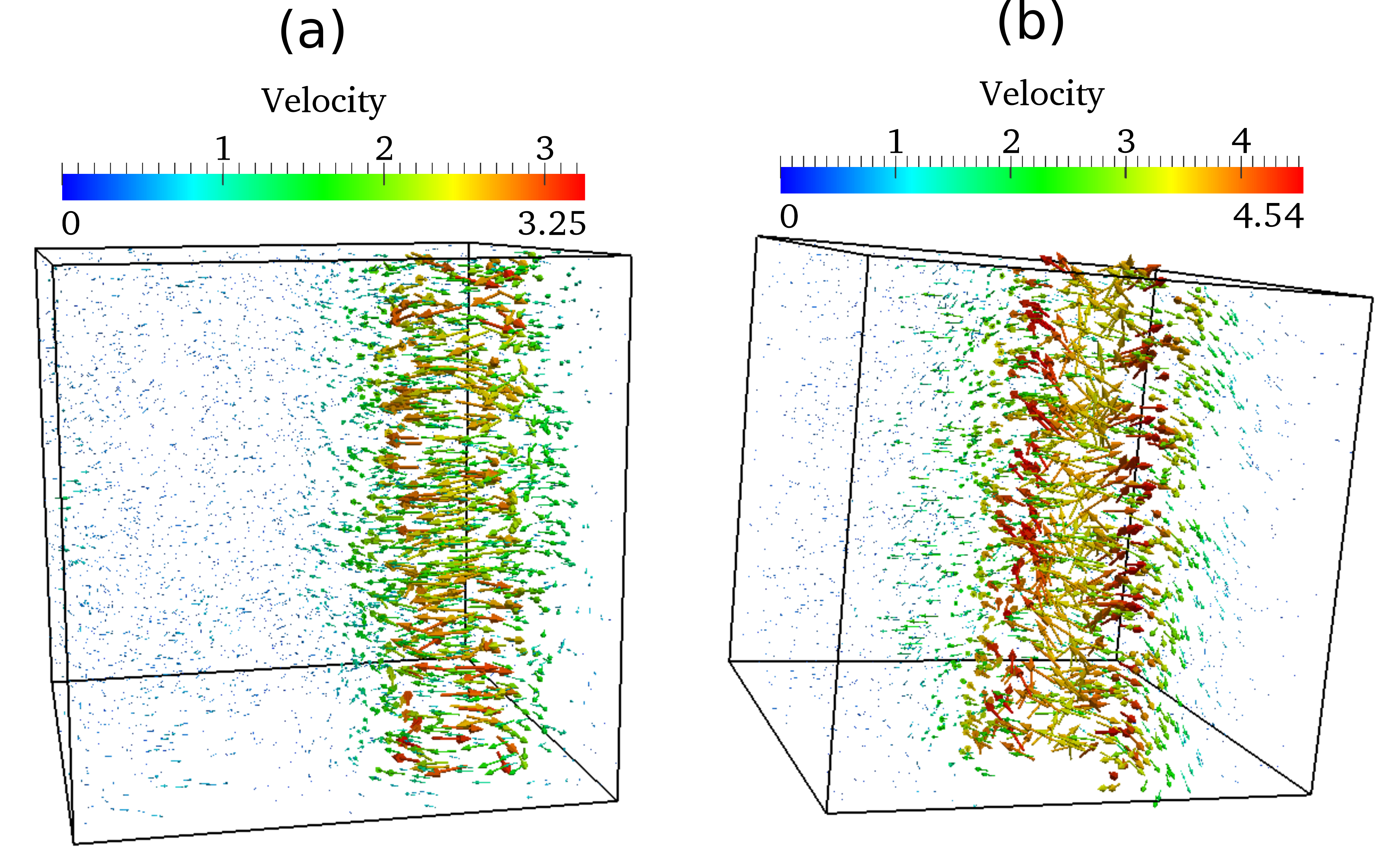}
 \end{center}
\caption{Vector plot of the velocity field for (a) $N=18$ and (b) $N = 130$.  Figure (a) is an example of 2D flow, while (b) an example of two-dimensional three-component (2D-3C) flow.}  
 \label{fig:N132_vectors}
 \end{figure}

\begin{figure}[htbp]
 \begin{center}
 \includegraphics[scale=0.26]{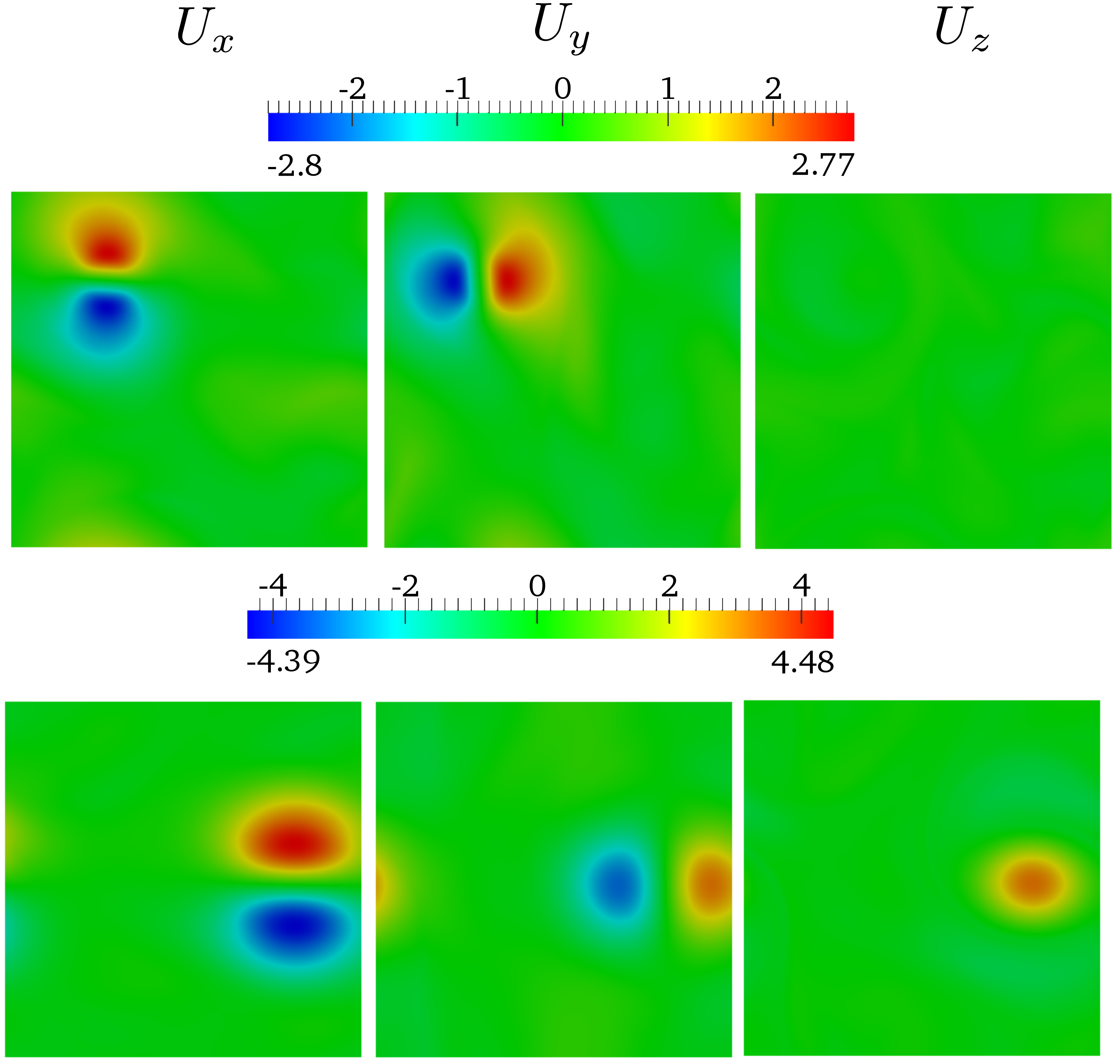}
 \end{center}
\caption{The magnitudes of $u_x$ (left), $u_y$ (middle), and $u_z$ (right) on a horizontal cross section at $z = 3.14$. Top row with $u_z \ll u_x,u_y$ is for $N = 18$, and bottom row with $u_z \sim u_x,u_y$ is for $N=130$.}  
 \label{fig:plane_N}
 \end{figure}

 In Fig.~\ref{fig:isosurface_vorticity} we exhibit isosurfaces of  the vorticity-field amplitudes for $N=0$, $5.5$, and $18$. The flow develops strong vortical structures as $N$ is increased.  The strong vortex tube for $N=18$ demonstrates an approximate two-dimensional nature of the flow.   A careful examination of the field configurations show that the field is two-dimensional for $N = 11-18$ with most of the energy residing in the horizontal components of the velocity (perpendicular to the mean magnetic field).  However, the parallel component of the velocity starts getting quite significant from $N=27$ onwards.   We contrast the two configurations in Fig.~\ref{fig:N132_vectors}, where we illustrate the vector plots of the velocity field for $N=18$ and $130$.  Along with these plots, we also exhibit the density plots of the three components $u_x$, $u_y$ and $u_z$ in Fig.~\ref{fig:plane_N}.  These figures indicate that the flow field for $N=18$ is approximately two-dimensional (2D) with $(|u_x|  \sim |u_y|) \gg |u_z|$.  But for $N=130$, $|u_z|$ is comparable to  $|u_x|$ and $|u_y|$,  but the flow field is approximately dependent on $x$ and $y$ coordinates.  Thus, the flow field for $N=130$ is an example of a two-dimensional three-component (2D-3C) flow.

The aforementioned results are in qualitative agreement with those of Favier \textit{et al.,}\cite{Favier:PF2010b} but they differ in detail.  Favier \textit{et al.}\cite{Favier:PF2010b} report 2D-3C flow behaviour for $N_0=5$ itself for their decaying simulation.  However our numerical results show that the transition from 2D to 2D-3C behaviour is near $N=27$ or $N_0=20$.  The difference is probably due to the forcing applied in our simulations.  Also, it is possible that the flow for $N_0=5$ could become approximately two-dimensional in the asymptotic limit at a later time. 

\begin{figure}[htbp]
\begin{center}
\includegraphics[scale=0.45]{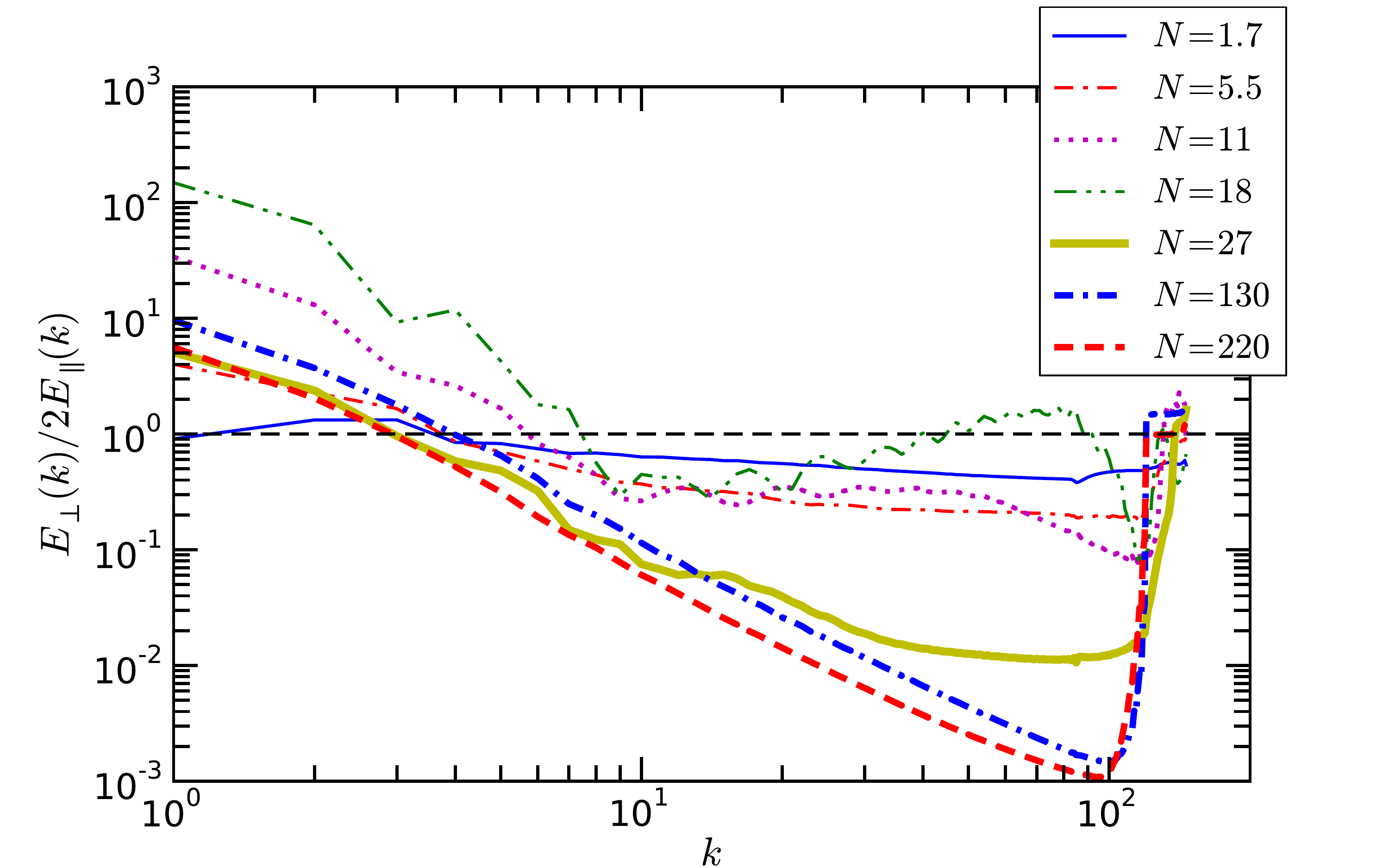}
\end{center}
\caption{Variation of $E_{\perp}/2E_{\parallel}$ with $k$ for different interaction parameters $N$.}  
\label{fig:Eperp_Epar}
\end{figure}

To explore the nature of anisotropy at different length scales, we study the wavenumber dependence of anisotropy $E_{\perp}(k)/2E_{\parallel}(k)$\cite{Vorobev:POF2005} (sum over the modes within a shell of radius $k$), and plot it in Fig.~\ref{fig:Eperp_Epar}.  The plot shows that $E_{\perp}(k) > E_{\parallel}(k)$  at low wavenumbers  (due to inverse cascade), while $E_{\parallel}(k)> E_{\perp}(k)$  at higher wavenumbers.\cite{Favier:PF2010b}  Interestingly, for large $k$, $E_{\perp}(k)/2E_{\parallel}(k)$  decreases monotonically with the increase of $N$.  For small $k$,  $E_{\perp}(k)/2E_{\parallel}(k)$ increases with $N$ up to $N=18$, after which it decreases.  These results are qualitatively similar to the Vorobev \textit{et al.,}\cite{Vorobev:POF2005} but our simulations have been carried out in more detail and for larger $N$.  These results are consistent with Favier \textit{et al.}'s\cite{Favier:PF2010b} arguments that the horizontal velocity field has inverse cascade thus enhancing $E_\perp(k)$ for small $k$, while the parallel component has a forward cascade that leads to an increase in $E_{\parallel}(k)$ for large $k$.   Thus, $u_z$ is significant in 2D-3C flows at small scales.  We will revisit these issues in Sec.~\ref{sec:spectrum of KE}.

In this section, the anisotropy of the flows has been described  by global energy and the shell spectrum, which do not provide information about the angular dependence of energy.  We discuss this issue in the next section.

\section{Angular Energy Spectrum and  Legendre Polynomials} \label{sec:Legendre}

For isotropic flows, the energy of all the modes in a thin wavenumber shell are statistically equal.  Hence, it is customary in turbulence literature to report one-dimensional energy spectrum, which is the sum of energy of all the modes in the shell.  However, an application of the magnetic field induces anisotropy leading to an unequal distribution of the energy for  various modes in a shell.  To quantify this anisotropy, we divide a given shell into rings, which are indexed using the shell index $n$ and sector index $\alpha$\cite{Teaca:PRE2009} (see Fig.~\ref{fig:ring_decomposition} for an illustration).  Note that the mean magnetic field is aligned along $\theta=0$.  We define the ring spectrum as 
\begin{equation}
E(k,\theta) = \frac{1}{C_{\alpha}} \sum_{k \le |{\bf k'}| < k+1; \mathrm{\angle}({\bf k^\prime})\in [\theta_\alpha,\theta_{\alpha+1})} \frac{1}{2} |{\mathbf U}({\mathbf k'})|^2,
\end{equation}
where $\mathrm{\angle}({\bf k^\prime})$ is the angle between ${\bf k^\prime}$ and ${\bf B_0}$, and  $\alpha$ is the index of the sector whose range of angles vary from $\theta_{\alpha}$ to $\theta_{\alpha+1}$, and
\begin{equation}
C_{\alpha} = |\cos(\theta_{\alpha}) - \cos(\theta_{\alpha+1})|
\end{equation}
is the normalization factor that compensates the effects of larger number of modes in the rings with larger $\theta$; recall the $d\cos\theta$ factor in the volume integral in spherical geometry.  $E(k,\theta) $ is essentially a measure of the normalized energy per mode  in the ring.

 \begin{figure}[htbp]
 \begin{center}
 \includegraphics[scale=0.50]{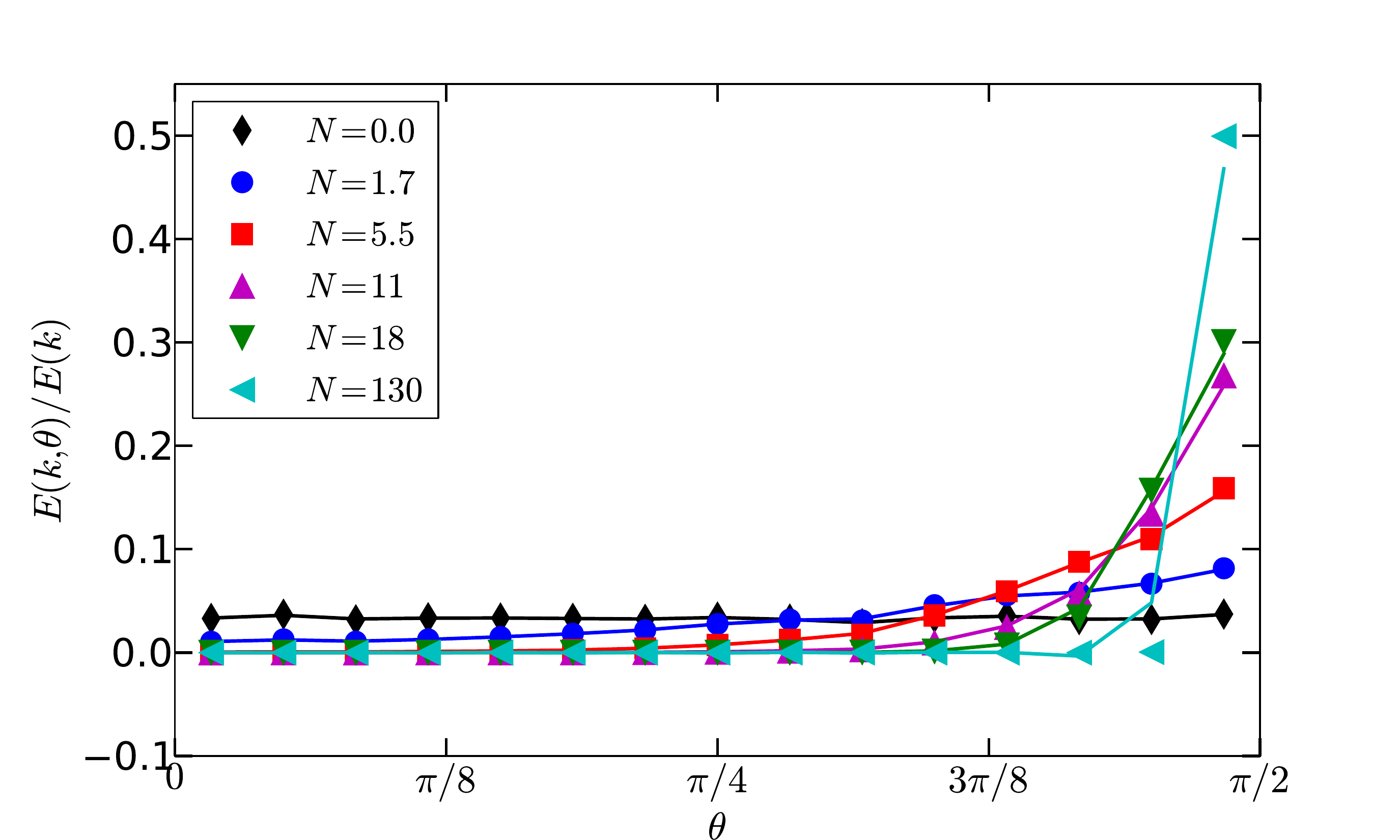}
 \end{center}
 \caption{Plot of $E(k=20,\theta)/E(k=20)$ vs. $\theta$. Markers represent simulation data, while the solid lines is  $E(k,\theta)$ computed using the polynomial expansion of Eq.~(\ref{eq_legen}). } 
 \label{fig:Ek_20}
\end{figure}

 In our simulations, we divide the spectral space in the northern hemisphere into thin shells, which are further divided into 15 thin rings from $\theta=0$ to $\theta=\pi/2$.  We do not compute the energy of the rings in the southern hemisphere due to $\theta \rightarrow \pi-\theta$ symmetry. Fig.~\ref{fig:Ek_20} exhibits the normalized ring spectra $E(k=20,\theta)/E(k=20)$ vs.~$\theta$ for $N=0$, 1.7, 5.5, 11, 18, and $130$.  The wavenumber $k=20$ is a generic wavenumber in the inertial range.  These plots show that for large $N$, the energy tends to be concentrated near $k_{||}=0$ or $\theta=\pi/2$ consistent with the experimental results of Caperan and Alemany,\cite{Caperan:JDM1985} and the numerical results of Burattini {\textit {et al.,}}\cite{Burattini:PD2008} and Potherat and Dymkou.\cite{Potherat:JFM2010}

\begin{figure}[htbp]
\begin{center}
\includegraphics[scale=0.46]{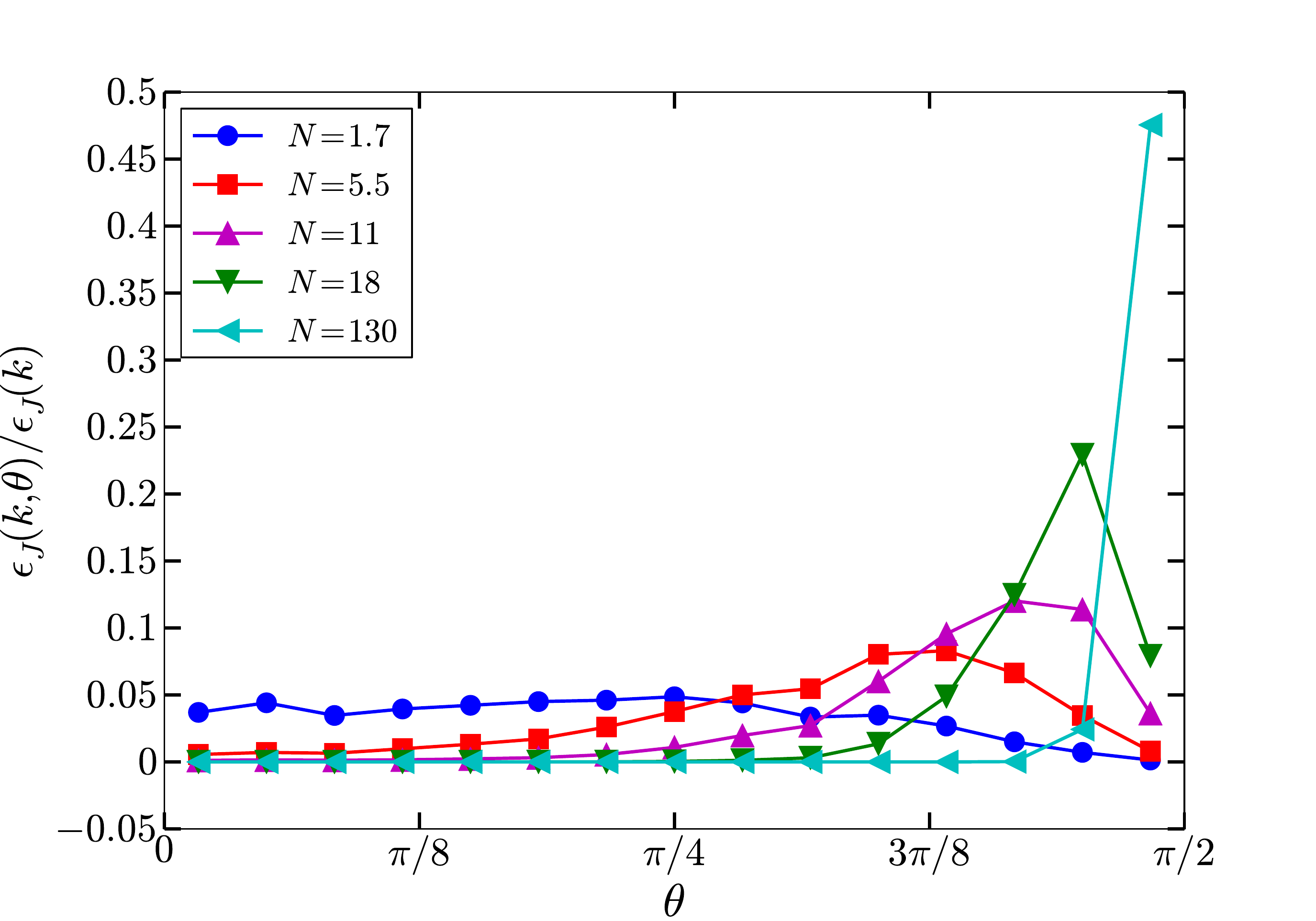}
\end{center}
\caption{Variation of normalized Joule dissipation rate $\epsilon_J(k=20,\theta)/\epsilon_J(k=20)$ for various interaction parameters $N$.}  
\label{fig:dissipation_k_20}
\end{figure}

The spectrum of viscous dissipation rate $\epsilon_\nu(k,\theta) = 2 \nu k^2 E(k,\theta)$ has similar angular distribution since $\epsilon_\nu(k,\theta) \propto E(k,\theta)$.  The angular distribution of the Joule dissipation rate however has an additional $\cos^2 \theta$ dependence: 
\begin{equation}
\epsilon_J (k,\theta) = 2{B'_0}^2 E(k,\theta)\cos^2\theta.
\end{equation}
The spectral energy density $E(k,\theta)$ is maximum near $\theta=\pi/2$, but $\cos^2\theta$ is minimum for this angle.  Hence, the product $E(k,\theta)\cos^2\theta$ peaks at an angle $\theta < \pi/2$.  Fig.~\ref{fig:dissipation_k_20} shows a plot of normalized Joule dissipation rate $\epsilon_J(k=20,\theta)/\epsilon_J(k=20)$ vs. $\theta$ for $N=0$, 1.7, 5.5, 11, 18, and $130$. The plots show that the maximum value of  $\epsilon_J (k,\theta) $  occurs near $\theta=\pi/2$ but not at $\pi/2$, consistent with our above arguments.  The normalized $\epsilon_J$ peaks near the equator, with its maxima shifting towards the equator with the increase in $N$; however, it vanishes at the equator.  This feature is absent for $N=130$, which is due to an insufficient angular resolution used in that simulation.  A computation of ring spectrum for $N=130$ requires more refinement near the equator, which is quite expensive.

\begin{figure}[htbp]
\begin{center}
\includegraphics[scale=0.47]{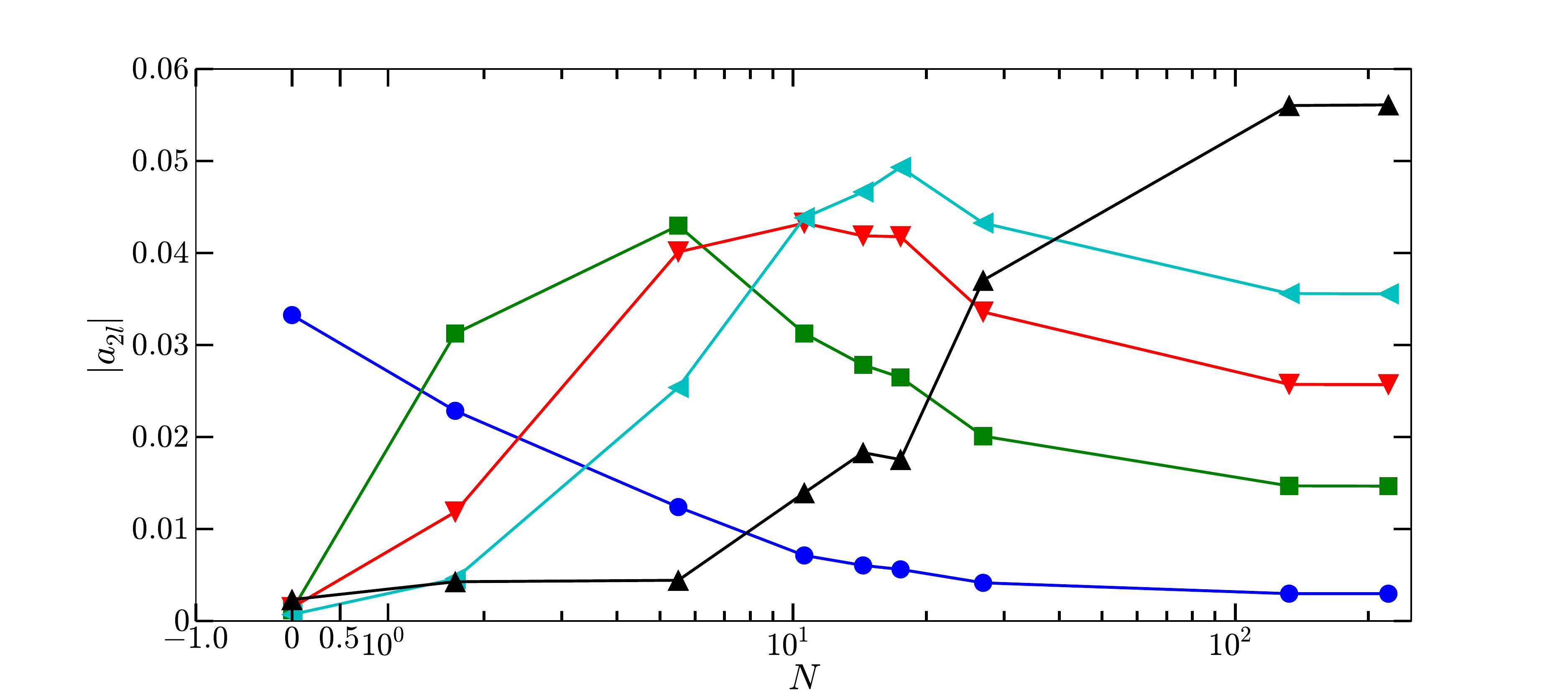}
\end{center}
\caption{The coefficients $a_l$ of Legendre polynomials  computed using the numerical data for $N = 0-220$. Here, $a_0$ is represented by {\color{blue}$\newmoon$}, $a_2$ by {\color{ForestGreen}$\blacksquare$}, $a_4$ by {\color{red}$\blacktriangledown$}, $a_6$ by {\color{Turquoise}$\blacktriangleleft$}, and $a_{16}$ by $\blacktriangle$. }
\label{fig:Legen_coeff}
\end{figure}

  The above description of anisotropy is qualitative.  We quantify the measure of anisotropy using spherical harmonics, which is a preferred basis function based on group-theoretic arguments.\cite{Biferale:PR2005}  In particular, we use Legendre polynomials to extract angular dependence of the large scale flow.  This is in a similar spirit as the ``proper orthogonal decomposition" or ``mode analysis".\cite{Chandra:PRE2011}    In terms of physical interpretation, the energy of isotropic flows are constant in polar angle $\theta$, hence it can be described by the zeroth component of the Legendre polynomial.  An introduction of external field, e.g., magnetic field, makes the energy spectrum ($E(k,\theta)$) a function of $\theta$.  The Legendre polynomials are convenient description of the anisotropic angular dependence of the spectrum.  Higher components of the Legendre polynomials become important for strongly anisotropy flows.

For the liquid metal flows discussed in this paper,   the energy and dissipation spectra are independent of the azimuthal angle $\phi$ due to azimuthal symmetry of the system.  Therefore,  $E(k,\theta)$ can be expanded as
\begin{equation}
E(k, \theta) = \sum_l a_l P_{l}(\cos \zeta), \label{eq_legen}
\end{equation} 
where the angle $\zeta=\pi/2-\theta$ is chosen so as to keep the maximum of the function for $\zeta=0$.  We compute the coefficients $a_l$ using our numerical data.  Note that the computation of $a_l$ requires data for $\zeta=[0,\pi]$; for $\zeta=[\pi/2,\pi]$ we use the $\theta \rightarrow \pi-\theta$ symmetry.  We use $l=0-28$ for our expansion.  We observe that the odd $a_l$'s are negligible due to the $\theta \rightarrow \pi-\theta$ symmetry.  Fig.~\ref{fig:Legen_coeff} exhibits some of the generic even $a_l$ coefficients.  For $N=0$, $a_0$ is much larger than the other coefficients, which is consistent with the isotropic nature of the flow for $N=0$.  For larger $N$'s,  $a_0$ decreases and higher $a_l$'s ($l>0$) become significant.  For the coefficients shown in the figure, $a_2$ and $a_4$ are most dominant for $N=5.5$, while $a_6$ and $a_{16}$ dominate for $N=18$ and $N=220$ respectively.   We observe that the magnitudes of the  higher order Legendre modes increase with the increase of $N$, thus signaling stronger anisotropy for larger $N$.   For $N=220$, the $16$th Legendre mode is most dominant, which indicates that most energy is concentrated near the equator for this parameter.  This is consistent with our numerical observations as well as earlier results.~\cite{Burattini:PD2008,Caperan:JDM1985,Potherat:JFM2010}

In the next section we will discuss the energy spectrum and energy flux for large $N$ simulations.


\section{Kinetic energy spectrum} \label{sec:spectrum of KE}
Energy spectrum for classical hydrodynamic turbulence is described by  Kolmogorov's theory as\cite{Kolmogorov:DANS1941a}
\begin{equation}
E(k) = K_{Ko} (\Pi(k))^{2/3} k^{-5/3}
\label{eq:kolm}
\end{equation}
where $K_{Ko}$ is Kolmogorov's constant, and $\Pi(k)$ is energy flux crossing the   spectral sphere of radius $k$.  In Kolmogorov's theory, the flux $\Pi(k)$ is assumed to be independent of $k$ since the viscous dissipation is effective only at very high $k$.   For higher $N$'s, Joule dissipation reduces the energy flux in each shell, which yields a wavenumber dependent $\Pi(k)$.\cite{Verma:ARXIV2013a}   However, this argument is applicable for small $N$.  For large $N$, the velocity field is two-dimensional with three components (2D-3C), whose energy spectrum is very different from that described by Eq.~(\ref{eq:kolm}).

\begin{figure}[htbp]
 \begin{center}
 \includegraphics[scale=0.4]{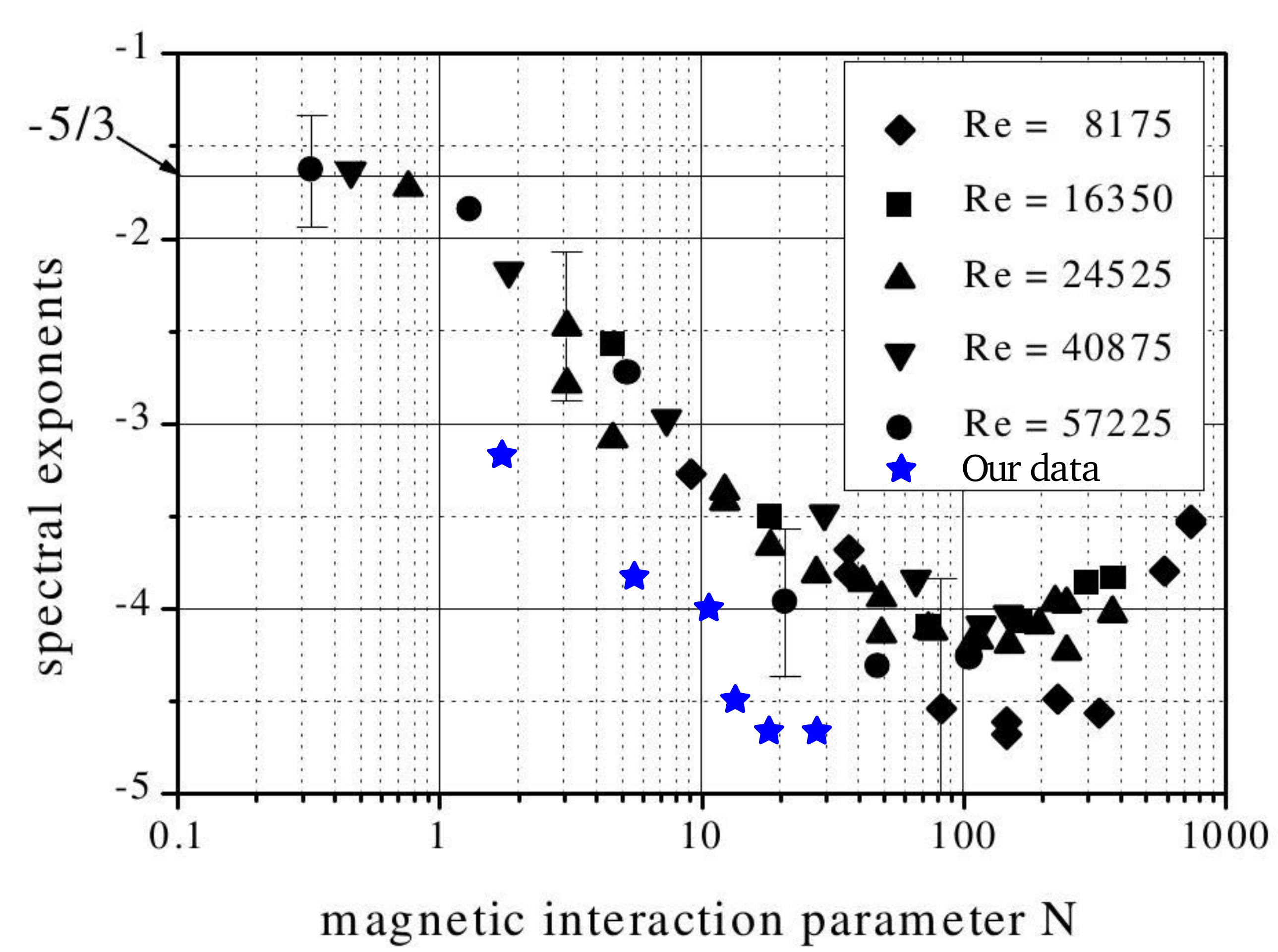}
 \end{center}
 \caption{Scaling of exponent of spectrum with $N$. Figure adopted\cite{permission} from Eckert \textit{et al.}\cite{Eckert:HFF2001}}  
 \label{fig:Eckert}
 \end{figure}

\begin{figure}[htbp]
 \begin{center}
 \includegraphics[scale=0.43]{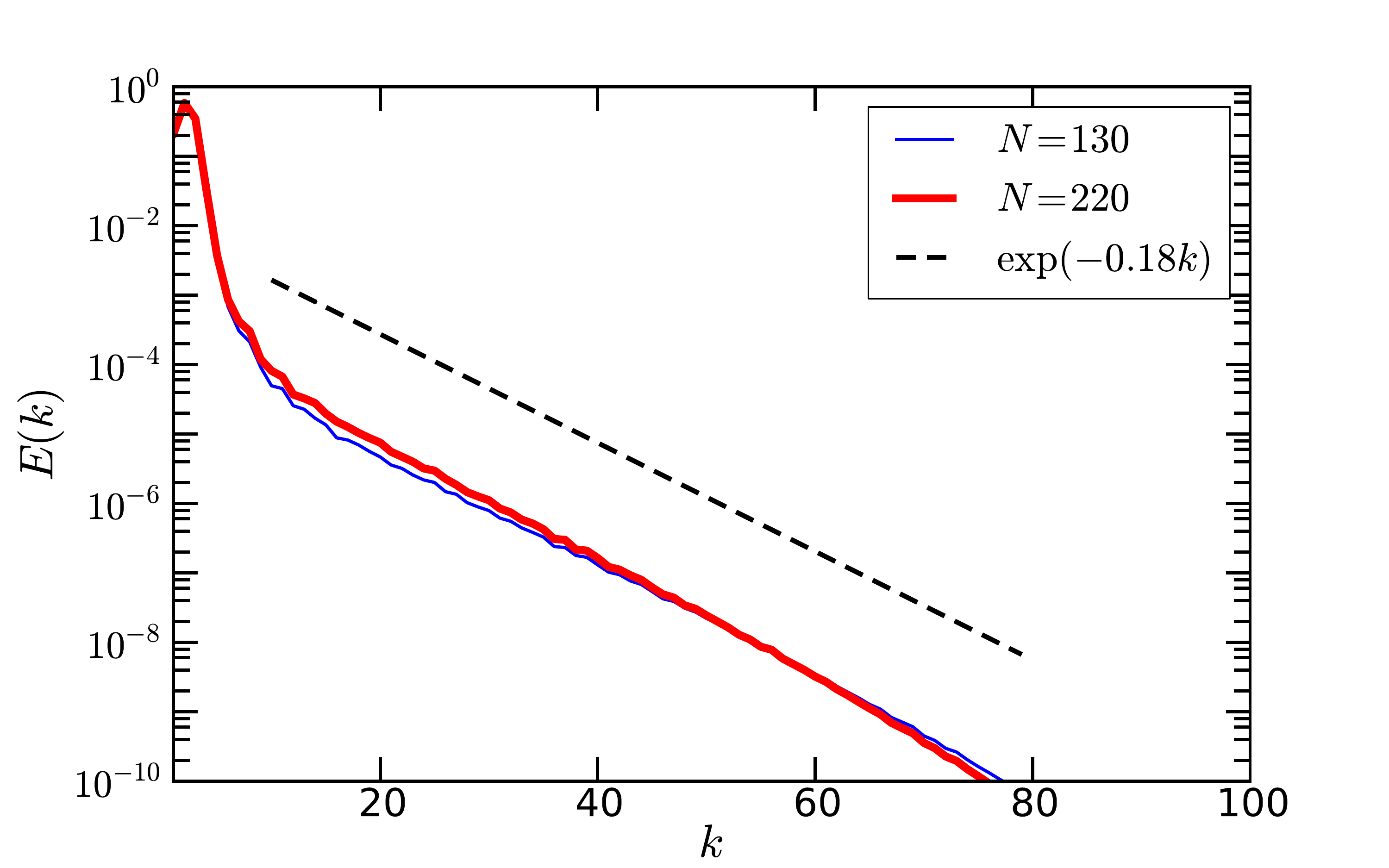}
 \end{center}
 \caption{Kinetic energy spectrum for $N=130$ and $N=220$, which are extreme  $N$'s. The dashed  line represents $\mathrm{exp}(-0.18k)$, thus demonstrating an exponential behaviour for very large $N$.}  
 \label{fig:KE_spectrum_semilog}
 \end{figure}
 
 \begin{figure}[htbp]
 \begin{center}
 \includegraphics[scale=0.6]{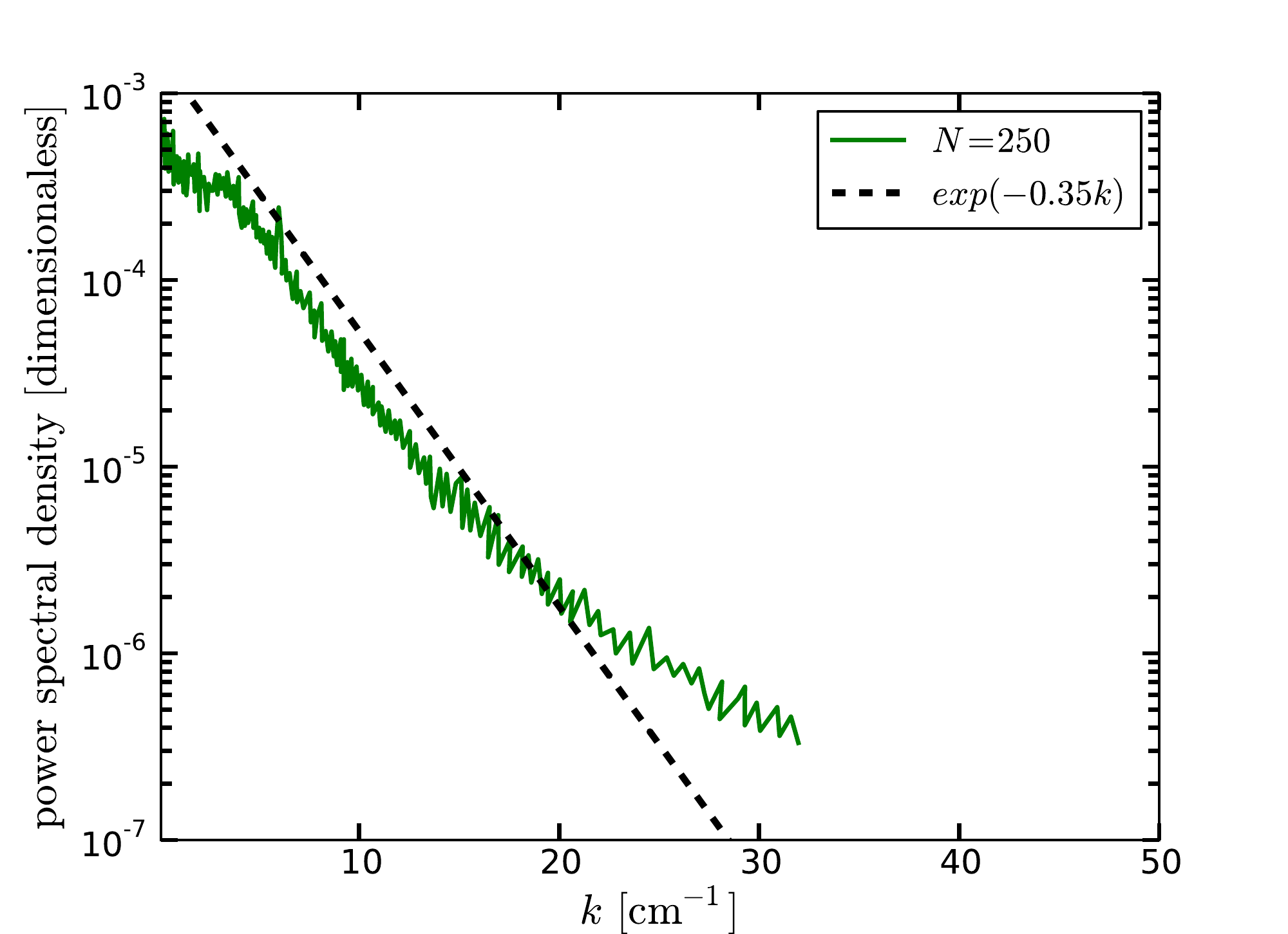}
 \end{center}
 \caption{Kinetic energy spectrum shown in semi-log scale for $N=250$ using digitized data of Fig.~9\cite{permission} in Eckert {\it et al.}\cite{Eckert:HFF2001}}  
 \label{fig:eckert_exp}
 \end{figure}

 We compute the energy spectrum under steady state for various $N$'s.  For $N=0$, which is   classical hydrodynamic simulation, we obtain Kolmogorov-spectrum.  For $N=1.7-27$, the energy spectrum is a power law, with spectral exponent  ranging from 3.2 to 4.7, which are exhibited as blue stars in  Fig.~\ref{fig:Eckert}, and in Table~\ref{tab:parameters}.  Our results are in qualitative agreement with the spectral exponents obtained by  Eckert \textit{et al.}\cite{Eckert:HFF2001} from their experimental data (also exhibited in Fig.~\ref{fig:Eckert}).  The difference between the two results is expected due to the absence of wall effects in our simulation. For very large $N$'s ($130$ and $220$), the energy spectrum obtained from our numerical data is exponential with $E(k) \sim \exp{(-0.18k)} $ (see Fig.~\ref{fig:KE_spectrum_semilog}). We also performed a similar analysis on the digitized data of Fig. 9 of  Eckert \textit{et al.}\cite{Eckert:HFF2001} and observed that an exponential function is a better fit than a power law function (see Fig.~\ref{fig:eckert_exp} and Fig.~9 of Eckert \textit{et al.}\cite{Eckert:HFF2001}), consistent with our numerical results.  The exponential energy spectrum is expected for very large $N$ flows due to a strong Joule dissipation in the flow; this result is similar to the exponential energy spectrum observed for laminar flows for which the nonlinearity is very weak. Note that the arguments of the exponential function for the numerical result ($-0.18 k$) and the experimental result ($-0.35k$) are somewhat different. This may be because our analysis is for a periodic boundary condition, while the experimental and realistic flows have no-slip boundary condition for the velocity field.  Different definitions of the interaction parameter may also play a factor for the different exponential functions.

Branover \textit{et al.}\cite{Branover:book}  attribute the steepening of the energy spectrum of the liquid metal flows to the helicity in the flows, while Verma\cite{Verma:ARXIV2013a}  attempts to explain this phenomenon using variable energy flux. The Joule dissipation in quasi-static MHD flows is active at all scales unlike viscous forces  which are dominant in the dissipation range. For $N=0$ (hydrodynamic flows), from Kolmogorov theory, energy flux $\Pi(k)$ is a constant in the inertial range. For $N\neq 0$, the presence of Joule dissipation acting at all scales leads to a decrease of $\Pi(k)$ with $k$ in the inertial range itself.  A substitution of such $k$-dependent $\Pi(k)$ in Kolmogorov's formula  $E(k) \propto (\Pi(k))^{2/3} k^{-5/3}$ yields a lower spectral exponent than -5/3. For large $N$, the flow is dominated by Joule dissipation, which steepens the spectrum further to an exponential form.  This is similar to the exponential energy spectrum for laminar flows ($\mathrm{Re} \lesssim 1$), as well as for two-dimensional flows with strong Ekman damping.\cite{Verma:EPL2012}

\begin{figure}[htbp]
\begin{center}
\includegraphics[scale=0.22]{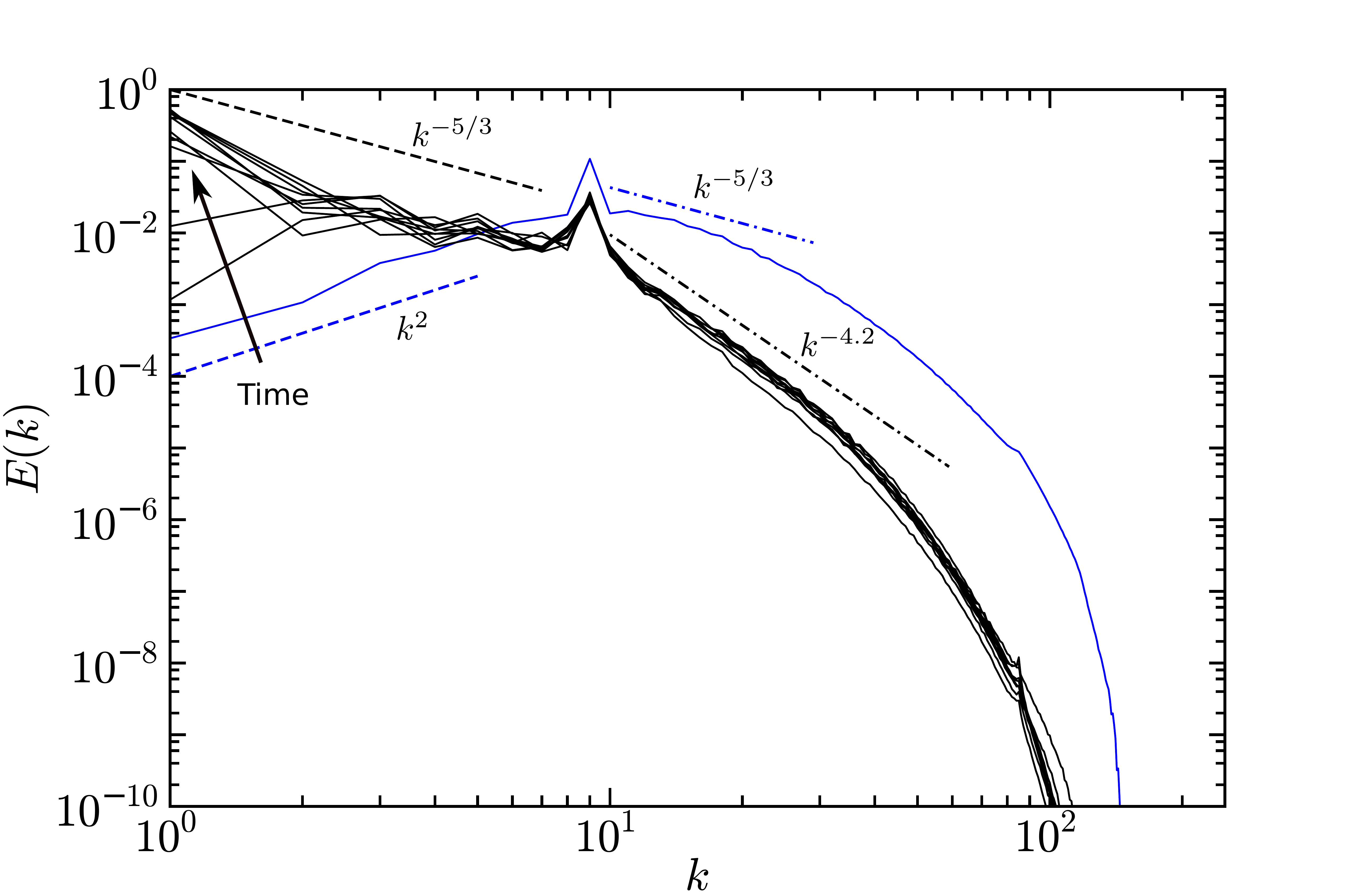}
\end{center}
\caption{Time evolution of energy spectrum for forcing with $k_f = [8,9]$. Solid blue line represents spectrum of fluid simulation ($N = 0$). Solid black lines represents energy spectrum at different times for $N = 100$.
The asymptotic curve shows $k^{-5/3}$  energy spectrum for low wavenumbers, thus indicating an inverse cascade in this regime. }  
\label{fig:2Dspectrum}
\end{figure}

Aforementioned discussion and earlier work indicate that two-dimensionality plays an important role in quasi-static MHD turbulence.  However, the energy flux for the flow has not been investigated in detail.  In the following discussion we compute energy spectrum and energy flux for $N=100$ ($N_0=30$).  Our simulations discussed so far had forcing band $k_f = [1,3]$, which prohibits a detailed investigation of inverse cascade.   Note that $k_\mathrm{min}=1$ in our simulations. To explore a possibility of an inverse cascade,  we study the energy flux in a new set of simulations for a forcing band of $k_f = [8,9]$.

\begin{figure}[htbp]
\begin{center}
\includegraphics[scale=0.25]{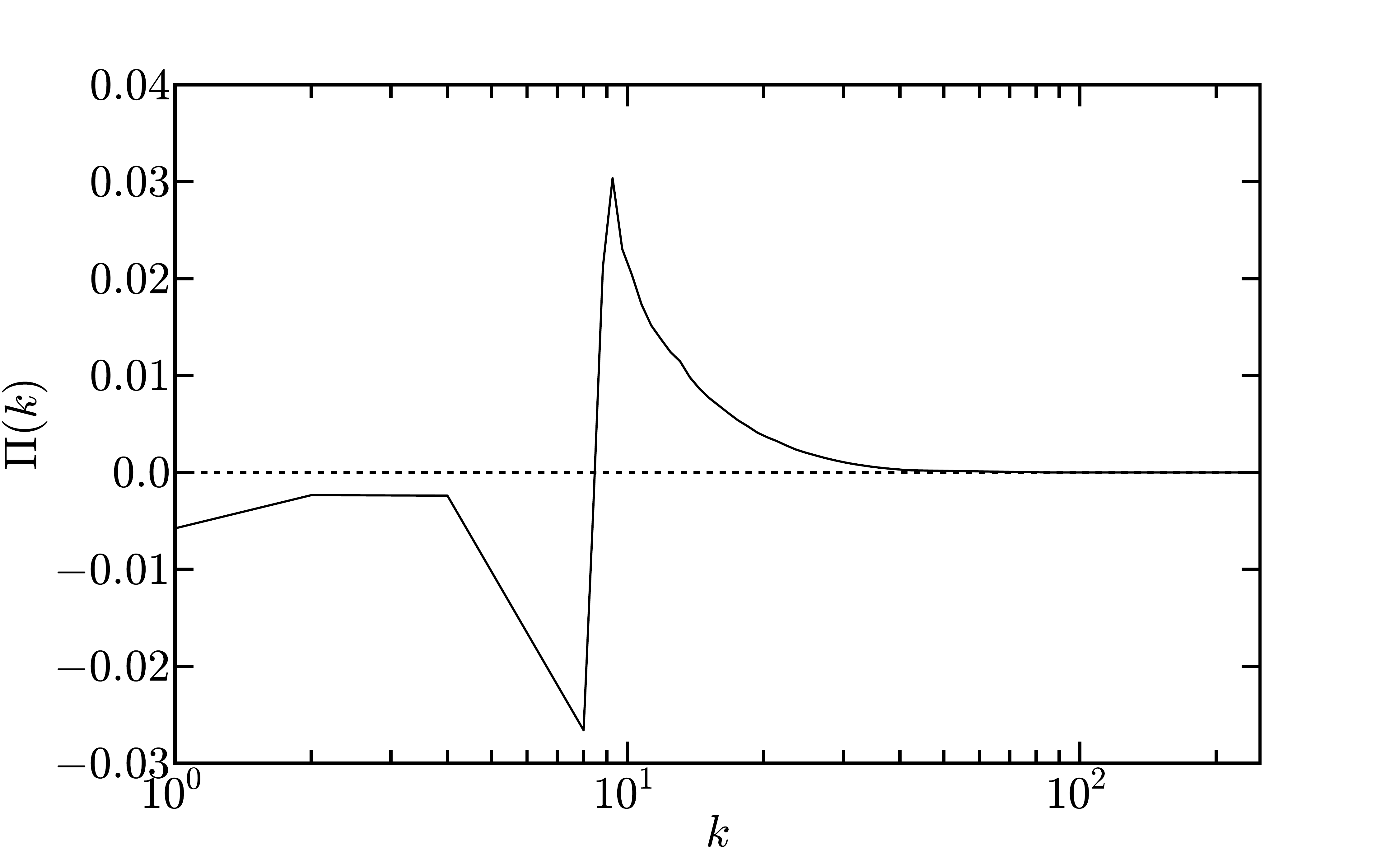}
\end{center}
\caption{Energy flux for $N = 100$ with forcing applied to wavenumbers in the shell $k_f = [8,9]$.  The figure exhibits an inverse cascade  of energy flux at low wavenumbers.}  
\label{fig:flux}
\end{figure}

For the new run we apply the same forcing scheme (except for the shifted wavenumber band) and initial condition as before (see Sec.~3).  First, the system is evolved for $N=0$ with the aforementioned forcing till a steady state is reached.  We observe a narrow $k^{-5/3}$ energy spectrum in the inertial range, and $k^{2}$ spectrum for the low wavenumber modes.   We take the final state of $N=0$ run, and then use it as an initial condition for a simulation with  $N=100$ ($N_0=30$), and evolve the flow until it reaches a new steady state.  Under steady state, the flow exhibits $k^{-5/3}$ energy spectrum for $k<k_f$ and $k^{-4.2}$ for $k>k_f$ (see Fig.~\ref{fig:2Dspectrum}).  In addition, we compute the energy flux, which is plotted in Fig.~\ref{fig:flux}.  The figure exhibits negative energy flux for $k<k_f$. Note however that the energy spectrum for $k>k_f$ is steeper than $k^{-3}$, which is due to the aforementioned variable energy flux caused by the action of the Joule dissipation at all scales.

It is interesting to note that the nonlinear energy flux appears to play an important role even for very large $N$, for which the flow is essentially laminar.  This is because the Lorentz force $N \cos\theta$  becomes negligible near the equatorial plane, and the nonlinear term dominates the dynamics near the equatorial plane for very large $N$.

\section{Discussions and Conclusions} \label{sec:conclusion}

In this paper we study various properties of quasi-static MHD turbulence for large interaction parameters ($N$).  Our maximum $N$ is $220$, which is much larger than those investigated by earlier researchers.   We employ direct numerical simulation with forcing.  It is important to note that the forced simulations have certain dissimilarities with decaying ones.

Main results of our simulations are as follows:

\begin{enumerate}

\item The external magnetic field induces anisotropy, which is quantified using $E_\perp/(2E_{||})$.  The ratio  increases all the way up to $\approx 16$ for $N=18$, after which it decreases non-monotonically.   The numerical values of $E_\perp/(2E_{||})$ observed in our simulations is much larger than those reported by Favier \textit{et al.}\cite{Favier:PF2010b} for decaying simulations with the same range of $N$.  The discrepancy is due to the forcing employed in our simulation.

\item We compute ring spectrum $E(k,\theta)$ that provides information about the angular distribution of energy.  We observe that the energy and viscous dissipation peak at the equator, but Joule dissipation is maximum near the equator, but not at the equator ($\theta=\pi/2$).  This shift is due to the $\cos^2\theta$ factor that vanishes for $\theta=\pi/2$.

We quantify the anisotropy by expanding the ring spectrum using Legendre polynomials, i.e., $E(k,\theta) = \sum_l a_l P_l(\cos (\pi/2-\theta))$.  We observe that $a_0$ is maximum for $N=0$, but the higher order $a_l$'s become prominent for larger $N$. The increase of prominent $l$ with $N$ is monotonic.

\item A careful observation of the numerical data reveals that the flow field is two-dimensional till $N$ up to 20 or so. For $N \ge 27$, the vertical component of the velocity field is  comparable to the horizontal components, which indicates two-dimensional three-components (2D-3C) type flow, reported by Favier \textit{et al.}\cite{Favier:PF2010b}  We observe that for all $N$, $E_\perp(k) \gg E_{||}(k) $ for small wavenumbers due to an inverse cascade of $u_\perp^2$.  However, $ E_{||}(k) \gg E_\perp(k)$  for large $k$ (see Fig.~\ref{fig:Eperp_Epar});  Favier \textit{et al.}\cite{Favier:PF2010b} attribute this strengthening of $ E_{||}(k)$ to a forward cascade of $u_z^2$.   Note that the change-over from 2D to 2D-3C behaviour occurs much earlier in Favier \textit{et al.}'s\cite{Favier:PF2010b} simulation, which may be due to the absence of external forcing in their simulation.

\item The shell spectrum $E(k)$ is a power law for moderate $N$ ($N \le 27$), with the spectral index  ranging from 3.2 to 4.7.  However, the spectrum becomes exponential for very large $N$.  The steepening of the spectrum is due to the combined effects of Joule dissipation and variable energy flux, inverse cascade of $u_\perp^2$, and forward cascade of $u_z^2$. This issue needs to be investigated in detail.

\item We compute the energy flux using the numerical data.  For forcing under a narrow band near $k_f  = [8,9]$,  we observe an inverse cascade of energy and $k^{-5/3}$ energy spectrum.  This is the first quantitative and direct computation of the inverse energy cascade  for the quasi-static MHD in the low wavenumber regime.  This feature is similar to the inverse cascade of energy observed in 2D fluid turbulence.  A similar quantitative computation of the forward cascade of the parallel velocity component would be very useful for understanding the significant buildup of $E_{||}(k)$ for large wavenumbers.
\end{enumerate}

In summary, our numerical simulations show some interesting properties of  quasi-static MHD under large $N$ limit.  For these cases, the energy spectrum is exponential, yet the energy flux is very significant.  A more refined description of energy flux with individual computations for $u_\perp^2$ and $u_z^2$ fluxes would be very useful for understanding various aspects of dynamics.  Favier \textit{et al.}\cite{Favier:PF2010b} bring out some interesting arguments for this regime with moderate $N$; these computations and arguments need to be extended to large $N$ cases.

\begin{acknowledgments}
We are grateful to the anonymous referees for valuable suggestions and comments. We thank Raghwendra Kumar, P. Satyamurthy, Prasad Perlekar for fruitful discussions, and Ambrish Pandey for tips on {\em matplotlib}. Simulations were performed on HPC system  and CHAOS cluster of IIT Kanpur. This project was supported by the research grant 2009/36/81-BRNS from Bhabha Atomic Research Center and Swarnajayanti fellowship from Department of Science and Technology, India.
\end{acknowledgments}

%
\end{document}